\documentclass[journal,comsoc]{IEEEtran}
\usepackage[T1]{fontenc}% optional T1 font encoding

\ifCLASSINFOpdf
   \usepackage[pdftex]{graphicx}
\else
\fi
\usepackage{amsmath}
\usepackage[cmintegrals]{newtxmath}
\usepackage{color}
\usepackage[square, comma, sort&compress, numbers]{natbib}
\usepackage{makecell}

\usepackage{algorithm}
\usepackage{algpseudocode}
\usepackage{amsmath}

\ifCLASSOPTIONcompsoc
 \usepackage[caption=false,font=normalsize,labelfont=sf,textfont=sf]{subfig}
\else
 \usepackage[caption=false,font=footnotesize]{subfig}
\fi

\usepackage{booktabs}
\usepackage{multirow}

\usepackage{url}
\usepackage{dcolumn}
\begin{document}
\title{ Low Complexity Trellis-Coded Quantization in Versatile Video Coding}
\author{Meng~Wang, Shiqi~Wang,~\IEEEmembership{Member, IEEE,} 
        Junru~Li, Li~Zhang,~\IEEEmembership{Member, IEEE,} \\
        Yue Wang, Siwei~Ma,~\IEEEmembership{Member, IEEE} and Sam~Kwong,~\IEEEmembership{Fellow, IEEE}\\[0.5em]        
\thanks{
M. Wang, S. Wang and S. Kwong are with Department of Computer Science, City University of Hong Kong, Hong Kong, China, (e-mail: mwang98-c@my.cityu.edu.hk; shiqwang@cityu.edu.hk; cssamk@cityu.edu.hk). (\textit{Corresponding author: Shiqi Wang})

J. Li and S. Ma are with the Institute of Digital Media, Peking University, Beijing, China, (e-mail: junru.li@pku.edu.cn; swma@pku.edu.cn).

L. Zhang is with the Bytedance Inc., San Diego CA. USA, (e-mail: lizhang.idm@bytedance.com).

Y. Wang is with the Bytedance (HK) Limited., Hong Kong, China, (e-mail: wangyue.v@bytedance.com).
}
}
\maketitle

\begin{abstract}
The forthcoming Versatile Video Coding (VVC) standard adopts the trellis-coded quantization, which leverages the delicate trellis graph to map the quantization candidates within one block into the optimal path. Despite the high compression efficiency, the complex trellis search with soft decision quantization may hinder the applications due to high complexity and low throughput capacity.  
To reduce the complexity, in this paper, we propose a low complexity trellis-coded quantization scheme in a scientifically
sound way with theoretical modeling of the rate and distortion. As such, the trellis departure point can be adaptively adjusted, and unnecessarily visited branches are accordingly pruned, leading to the shrink of total trellis stages and simplification of transition branches.
Extensive experimental results on the VVC test model show that the proposed scheme is effective in reducing the encoding complexity by 11\% and 5\% with all intra and random access configurations, respectively, at the cost of only 0.11\% and 0.05\% BD-Rate increase. Meanwhile, on average 24\% and 27\% quantization time savings can be achieved under all intra and random access configurations. Due to the excellent performance, the VVC test model has adopted one implementation of the proposed scheme.
\end{abstract}


% \fi
\IEEEpeerreviewmaketitle

\section{Introduction}
Recent years have witnessed the rapid development of video coding technologies. Newly adopted coding tools afford more mode options to cope with various characteristics in video sequences, leading to significant improvement of coding efficiency. Video coding standards such as H.264/AVC~\cite{h264}, HEVC~\cite{Sullivan2013Overview}, AVS~\cite{avs2Ma} and VVC~\cite{vvc4}, specify the semantic of decoding process, bestowing space for encoder optimization and complexity reduction. To coordinate the behaviors of individual or multiple coding tools, rate distortion optimization (RDO)~\cite{RDO_sullivan} is employed all over the encoding stages. Consequently, the optimal combination of encoding modes and parameters can be systematically determined by RDO, resulting in further promotion of compression performance. Given the maximum allowed rate $R_{max}$, the aim of RDO is to minimize the encoding distortions by attempting different combinations of encoding parameters and modes from the set $\mathcal{M}$, which can be described as follows,
\begin{align}
    \min_{\mathcal{M}}\{D\} \quad \text{subject to } R \leq R_{max}.
    \label{MinD}
\end{align}
Herein, Lagrangian optimization~\cite{lambdaOpt} is used to convert the constrained problem in Eqn.~(\ref{MinD}) to an unconstrained one,
\begin{align}
    \min_{\mathcal{M}} \{J\}\quad \text{where } J = D + \lambda \cdot R,
\end{align}
where $J$ is the RD cost and $\lambda$ is the Lagrange multiplier. $R$ is the number of bits and $D$ indicates the distortion. In general, genuine encoding procedures such as transform, quantization, entropy coding, inverse quantization and inverse transform are obbligato to obtain $R$ and $D$ for an individual mode. Furthermore, numerous RD cost calculations shall be performed with different candidate modes and combinations. This extremely imposes heavy burdens in terms of computational complexity to the encoder, which may impede the implementation and applications of new video coding standards in real application scenarios.

Typically, there are three ways to economize the encoding computational complexity. 
The first achieves the bottom-level speedup by employing the single-instruction multiple-data (SIMD)~\cite{Ahn2014}, which focuses on specific modules that are with data-intensive operations, such as intra prediction, motion compensation, transformation and filtering. As a result, the operation time consumed by each mode attempting can be saved by performing SIMD without any performance variation. 
The second way alleviates the encoder burden based on pruning the mode attempting, which is able to reduce the number of RDO rounds. More specifically, improbable mode candidates are inferred experimentally or theoretically, and are directly ignored by skipping RD cost calculation or comparisons, leading to the savings of the overall encoding time~\cite{SHEN_FAST}. However, the remaining modes shall be evaluated with RDO.
The third way focuses on decreasing the complexity of RD calculation wherein the rate and distortion are estimated, instead of exhaustively going through the tedious working flow~\cite{HaibinYin,  FastRDOQ}.

The quantization, which is measured in terms of the goodness of the reproduced signal compared to the original as well as the resulting representation cost, evolves rapidly in video coding standards.
Soft decision quantization (SDQ) \cite{Trellis_H263,YangEnhui_2009, RDOQ_MARTA} introduces the sense of rate-distortion optimization to quantization level determination, which promotes the coding efficiency and simultaneously raises the computational complexity compared to the conventional hard decision quantization (HDQ)~\cite{HDQ}. During the standardization of H.264/AVC~\cite{h264}, HEVC~\cite{Sullivan2013Overview} and AVS2~\cite{avs2Ma}, a classical SDQ method, rate distortion optimized quantization (RDOQ)~\cite{RDOQ_MARTA}, was adopted and desirable coding performance had been achieved. However, the computational complexity of RDOQ becomes the barrier since the entropy coding shall be performed for each candidate along with context model updating. In VVC, besides RDOQ, trellis-coded quantization (TCQ)~\cite{Trellis_DCC} is adopted, which is also termed as dependent quantization. With TCQ, quantization candidates are delicately deployed into trellis graph at block level cooperating with state transfer, in an effort to convert the optimized quantization solution into the optimum path searching task. As such, the statistical dependencies among quantizaion outcomes within one coded block can be exploited. Different from the HDQ or RDOQ, where the former only conducts the quantization without considering the influence of coding bits, and the latter pays attention to the optimal RD behavior regarding the up-to-now coefficient, TCQ maps the coefficients to the trellis graph by employing vector quantizer, and seeks the path with the minimum RD cost as the optimal quantization solution. Superior coding performance is achieved by TCQ when compared to the RDOQ, where 3.5\% and 2.4\% bit-rate savings are reported under all intra (AI) and random access (RA) configurations, respectively, in the VTM-1.0 platform \cite{DQ_ICIP}. However, significant encoding complexity increase has also been observed, which attributes to the RD calculation, accumulation and comparison for each stage and each node during TCQ.

There have been many researches study on reducing the complexity of the sophisticated quantization process. Huang and Chen~\cite{RDOQ_HUANG} presented an analytical method to address the RDO-based quantization problem for H.264/AVC. In~\cite{FastRDOQ}, the variation of rate and distortion, $\Delta R$ and $\Delta D$ models were investigated, where improbable quantization candidates in RDOQ can be efficiently excluded. In~\cite{HybridLap}, transform coefficients are modeled with Laplacian distribution, which can be further utilized to deduce the block-level RD performance for RDOQ. A trellis-coded quantization method was studied in~\cite{YangEnhui_2009} for H.264/AVC, where all potential candidates along with coding contexts are mapped into the trellis graph, leading to the improvement of coding efficiency. However, the computational complexity is extremely high regarding the optimal path searching for both software and hardware implementations. To tackle this problem, Yin \textit{et al.}~\cite{HaibinYin} proposed a fast soft decision quantization algorithm that discriminated safe or unsafe quantization levels based on the speculation of the variations regarding rate and distortion.

Though the previous research works are effective for lessening the quantization complexities for H.264/AVC and HEVC, they are not applicable to the trellis-based quantization in VVC. Herein, we propose a low complexity TCQ scheme for VVC by modeling rate and distortion in a scientifically sound way. With the proposed model, the RD performance with regard to different quantization candidates can be effectively evaluated, and the computational intensive searching process in TCQ can be safely eliminated. In particular, the trellis departure point is adaptively determined, by which the total number of trellis stages can be shrunk. Moreover, the branch pruning scheme is investigated based on the RD models, which is conductive to decrease the operation complexity of TCQ.

\section{Statistical Rate and Distortion Models}
In the literature, the rate and distortion models have been statistically established according to the probability distribution of transform coefficients~\cite{LAPLACE_1983,DCT_QUANT_ERROR,LAM_DCT_MATH,Cauthy,XZhao2010, LiXiang}. The distribution of transform coefficients has been studied for several decades, including Laplacian distribution~\cite{LAM_DCT_MATH}~\cite{LiXiang}, Cauthy distribution~\cite{Cauthy}, generalized Gaussian distribution~\cite{XZhao2010} and combined distribution~\cite{TCM}. Generalized Gaussian distribution reveals the best modeling accuracy owing to the flexible controlling parameters associated to shape and scale. However, the controlling parameters are difficult to estimate, which significantly hinders its applications. In addition, Cauthy distribution may not be appropriate for the RD modeling task in the encoder since the mean and variance are not converged. By contrast, Laplacian distribution was regarded as the optimal solution for compromising modeling complexity and accuracy~\cite{LiXiang}.
In the literature, numerous rate and distortion models have been proposed, and they can be further applied to rate control, bit allocation and fast mode selection. A block level rate estimation scheme was presented in~\cite{XZhao2010} to speed up the RDO selection for H.264/AVC, where individual sub-band of transform coefficients is modeled with generalized Gaussian distributions. In~\cite{Cauthy}, frame-level bits are approximated and allocated by modeling the AC coefficients with Cauthy distribution. Moreover, the rate model was established from $\rho$-domain~\cite{Rho_domain} based on the assumption of Gaussian and Laplacian distribution, where a linear relationship between rate and the percentage of non-zero coefficients was delicately derived. In \cite{chang2003novel, Tu2006, RDOQ_HUANG, HEYUN_PCS}, the rate was also modeled with the $\ell_1$-norm of quantization coefficients.

In this section, we develop the rate and distortion models dedicated to the sophisticated designed quantization in VVC with Laplacian distribution.
In particular, let $C_s^{(i)}$ be the transform coefficient locating at position $i$ in a coding block with size $W \times H$, and the scalar quantization is given by,
\begin{equation}
    l^{(i)}_s = sign(C_s^{(i)})\cdot \left \lfloor\frac{|C_s^{(i)}|}{Q_{step}} + f \right \rfloor , i \in [0, W\times H - 1],
    \label{scalar_Quant}
\end{equation}
where $l^{(i)}_s$ is the corresponding scalar quantized coefficient. 
The parameter $f$ is typically involved to control the rounding offset, which is set to $1/2$ during the pre-quantization process in VVC and HEVC soft quantization. $Q_{step}$ represents the quantization step size.

Laplacian distribution is adopted here to model the transform residuals. In particular, the probability density function (PDF) of transform coefficient $C_s^{(i)}$ is given by,
\begin{align}
    p(x) = \frac{\Lambda}{2} e^{-\Lambda \cdot |x|},
    \label{LaplacePDF}
\end{align}
where $\Lambda$ is the Laplacian parameter that can be determined with the standard deviation $\sigma$ as follows,
\begin{align}
    \Lambda = \frac{\sqrt{2}}{\sigma}.
\end{align}

\subsection{Relationship between Rate and $\ell_0$-norm of Coefficients}
In general, given a certain allowed distortion level $D(x, \hat{x}) = |x-\hat{x}|$, based on Shannon's source coding theorem, the minimum bits for coding a symbol can be derived as,
\begin{align}
    R(D) = \log_2\left(\frac{1}{\Lambda \cdot D}\right).
    \label{rdLaplace}
\end{align}
As such, for a preset $Q_{step}$, the associated distortion is given by,
\begin{align}
    D(Q_{step}) = D_0(Q_{step}) + D_{\bar{0}}(Q_{step}),
\end{align}
where
\begin{align}
    D_0(Q_{step}) = 2 \cdot \int_0^{(1 - f) \cdot Q_{step}} p(x) \cdot x dx,
    \label{D_0}
\end{align}
\begin{align}
    D_{\bar{0}}(Q_{step}) = 2 \cdot \sum_{l=1}^{\infty}\int_{(l-f) \cdot Q_{step}}^{(l+1-f) \cdot Q_{step}}p(x) \cdot |x-l \cdot Q_{step}| dx.
    \label{D_non0}
\end{align}
Herein, $l$ represents the quantization level. As such, $D_0(Q_{step})$ and $D_{\bar{0}}(Q_{step})$ can be derived as follows,
\begin{align}
    &D_0(Q_{step}) = -\frac{1}{2}Q_{step}\cdot \tau + \frac{1}{\Lambda}\cdot(1- \tau),
\end{align}
\begin{align}
    D_{\bar{0}}(Q_{step}) &= \frac{1}{2}Q_{step}\cdot \tau
    + \frac{1}{\Lambda}\cdot\frac{\tau^{2}}{1-\tau^2}\cdot(2-\tau - \tau^{-1}),
\end{align}
where $\tau = e^{-\frac{1}{2}\Lambda Q_{step}}$. Therefore, $D(Q_{step})$ can be represented as,
\begin{align}
    D(Q_{step}) &= \frac{1}{\Lambda}\cdot \left[1 - \tau + \frac{\tau^2 \cdot(2-\tau - \tau^{-1})}{1-\tau^2}\right].
    \label{D_Q}
\end{align}
Given the PDF of the transform coefficients, the percentage of non-zero quantized coefficients can be estimated as follows,
\begin{align}
    P_{nz} &= 1 - \int^{(1-f)\cdot Q_{step}}_{-(1-f) \cdot Q_{step}}\frac{\Lambda}{2} e^{-\Lambda |x|}dx \nonumber \\
    &=\tau.
    \label{pnz}
\end{align}
Moreover, $P_{nz}$ can also be represented with $\ell_0$-norm,
\begin{align}
    P_{nz} = \frac{L_0}{W \times H},
\end{align}
where $P_{nz}$ should be within the range of [0,1] and $L_0$ denotes the $\ell_0$-norm in a coding block. Furthermore, the relationship between the coding bit and the percentage of non-zero quantized coefficients can be obtained by substituting Eqn.~(\ref{D_Q}) and Eqn.~(\ref{pnz}) into Eqn.~(\ref{rdLaplace}) as follows,
\begin{align}
    \hat{R_0} = \log_2\left(\frac{1+P_{nz}}{1-P_{nz}}\right). 
    \label{R_TAU}
\end{align}
The Taylor expansion of Eqn.~(\ref{R_TAU}) can be expressed as,
\begin{align}
    \hat{R_0} &= \frac{2}{\ln \left(2\right)}P_{nz}+\frac{2}{3\ln \left(2\right)}P_{nz}^3+\frac{2}{5\ln \left(2\right)}P_{nz}^5+\ldots.
    \label{taylorExp}
\end{align}
As such, we could have the following relationship,
\begin{align}
\hat{R_0} =& P_{nz}\cdot \left(\frac{2}{\ln \left(2\right)} +\frac{2}{3\ln \left(2\right)}P_{nz}^2+\frac{2}{5\ln \left(2\right)}P_{nz}^4+\ldots\right)\nonumber\\
\approx& \alpha \cdot L_0.
\label{taylorExp_1}
\end{align}
The relationship is approximated to be locally linear with respect to $P_{nz}$, which corresponds to the $\rho$-domain model~\cite{Rho_domain}.

%%%%
\subsection{Relationship between Rate and $\ell_1$-norm of Coefficients}
Herein, we further model the rate from the perspective of self-information~\cite{InformationTheory}. 
In particular, the self-information of a quantized symbol $l$ is given by,
\begin{align}
    \hat{r} = -\log_2p(l_s = l),
    \label{InfoTheory}
\end{align}
where $p(l_s = l)$ denotes the probability of the scalar quantization result $l_s$ equaling to $l$. Given the quantization step $Q_{step}$ and the rounding offset $f$, $p(l_s = l)$ is represented as follows,
\begin{equation}
    p(l_s = l)=
    \begin{cases}
        2 \cdot \int^{(1-f) \cdot Q_{step}}_{0}p(x)dx, & \text{$l = 0$}\\
        \int^{(|l|+1-f) \cdot Q_{step}}_{(|l|-f) \cdot Q_{step}} p(x)dx, & \text{$l\neq0$}.
    \end{cases}
    \label{p_L1}
\end{equation}
By integrating the Laplacian distribution into Eqn.~(\ref{p_L1}), the probability of the quantized symbol can be expressed as,
\begin{equation}
    p(l_s = l)=
    \begin{cases}
        1-e^{-\frac{1}{2}\Lambda Q_{step}} & \text{$l=0$}\\
        \frac{1}{2} \left(e^{-\Lambda Q_{step}(|l|-\frac{1}{2})} - e^{-\Lambda Q_{step}(|l|+\frac{1}{2})}\right) & \text{$l\neq0$}.
    \end{cases}
    \label{p_L1_solv}
\end{equation}
With Eqn.~(\ref{LaplacePDF}),  Eqn.~(\ref{InfoTheory}) and Eqn.~(\ref{p_L1_solv}), the rate of the quantized symbol can be approximated.
For the case of $l = 0$, $\hat{R_1}$ can be estimated as,
\begin{align}
    \hat{R_1} &= -\log_2(1-e^{-\frac{1}{2}\Lambda Q_{step}}) \nonumber\\ 
    &= \beta_0 \cdot |l| + b_0,
\end{align}
where
\begin{align}
    \beta_0 &= 0,\nonumber \\
    b_0 &= -\log_2\left(1 - e^{-\frac{1}{2}\Lambda Q_{step}}\right).
\end{align}
For the case of $l \neq 0$, $\hat{R_1}$ can be approximated as,
\begin{align}
    \hat{R_1} &= -\log_2\left[\frac{1}{2}\left( e^{-\Lambda Q_{step}(|l|-\frac{1}{2})} - e^{-\Lambda Q_{step}(|l|+\frac{1}{2})}\right) \right] \nonumber \\
    &= \Lambda Q_{step}\cdot\log_2(e)\cdot|l| + 1 - \log_2\left(e^{\frac{1}{2}\Lambda Q_{step}} - e^{-\frac{1}{2}\Lambda Q_{step}}\right) \nonumber \\
    &= \beta_1 \cdot |l| + b_1,
\end{align}
where
\begin{align}
    \beta_1 &= \Lambda Q_{step}\cdot\log_2(e), \nonumber \\
    b_1 &= 1 - \log_2\left(e^{\frac{1}{2}\Lambda Q_{step}} - e^{-\frac{1}{2}\Lambda Q_{step}}\right).
\end{align}
As such, the total coding bits of a coding block can be expressed as,
\begin{align}
    \hat{R_1} &= \beta_1 \cdot L_1 + b,
    \label{Eqn_L1_norm}
\end{align}
where $b$ represents the combination of $b_0$ and $b_1$.
In this regard, the number of coding bits of a block is determined by the $\ell_1$-norm of the coefficients.

\begin{figure}[t]
    \centering
    \includegraphics[width=8cm]{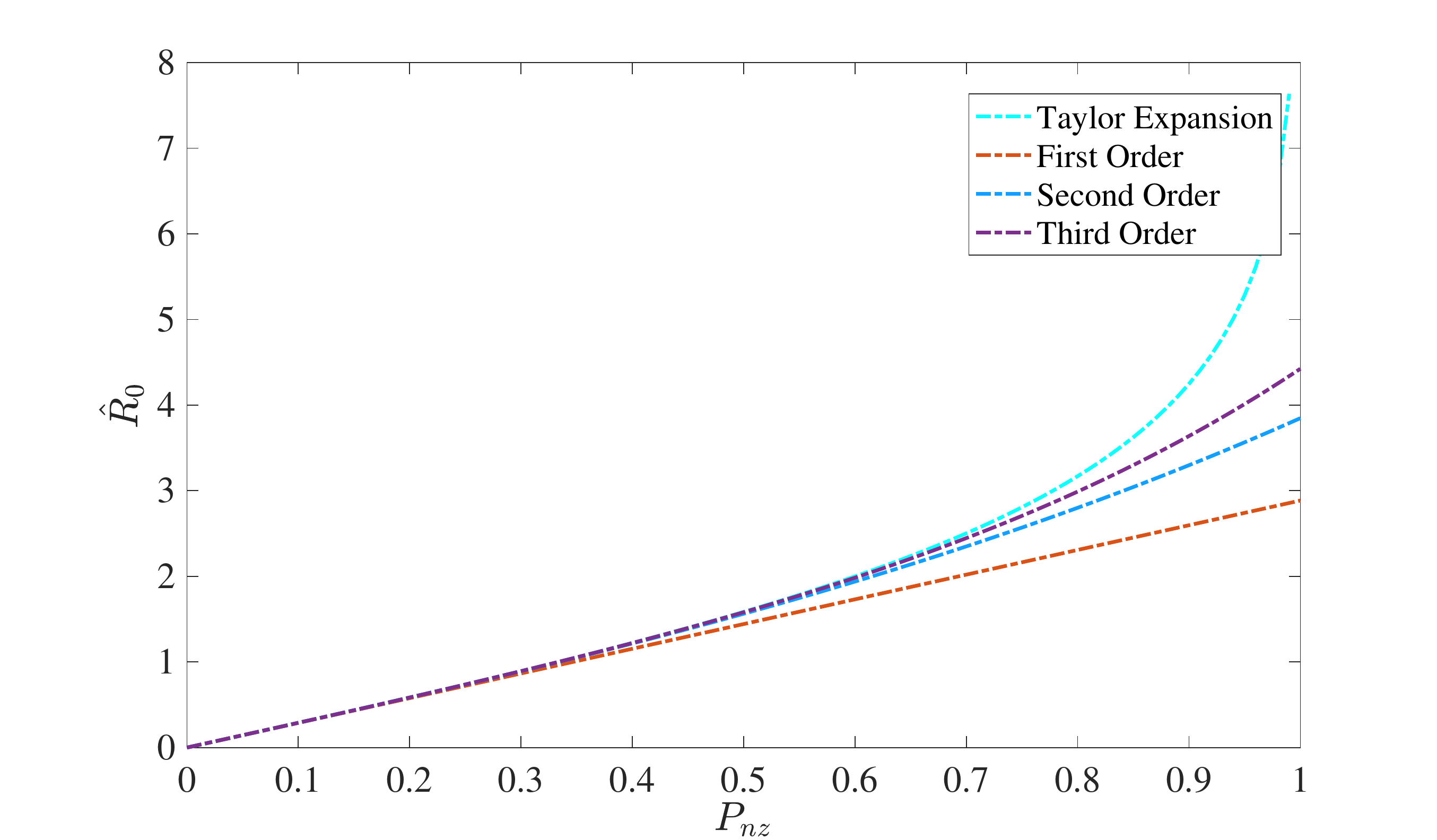}%
    \caption{Illustration of Taylor expansion, first order, second order and third order approximation with respect to $P_{nz}$.}
     \label{Fig_Taylor}
\end{figure}

\begin{figure}[!htb]
    \centering
    \subfloat[]{\includegraphics[width=0.87in]{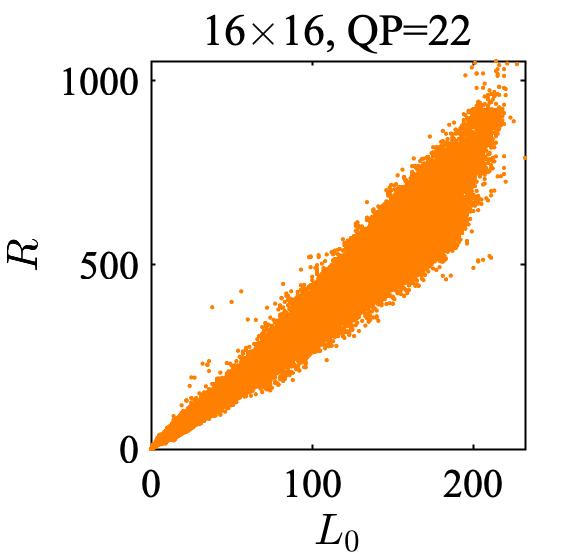}}
    \subfloat[]{\includegraphics[width=0.87in]{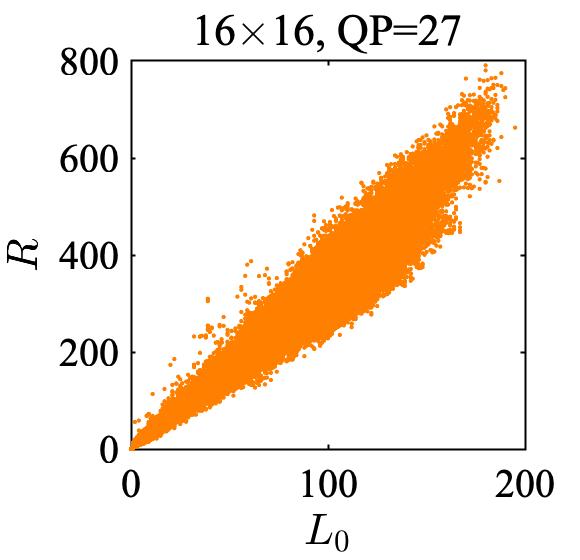}}
    \subfloat[]{\includegraphics[width=0.87in]{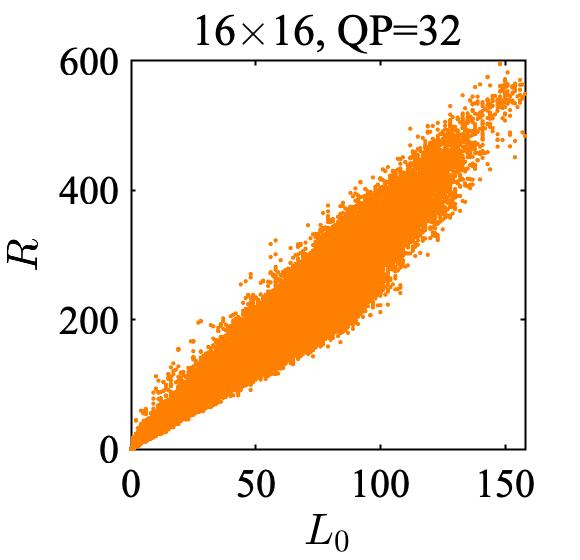}}
    \subfloat[]{\includegraphics[width=0.87in]{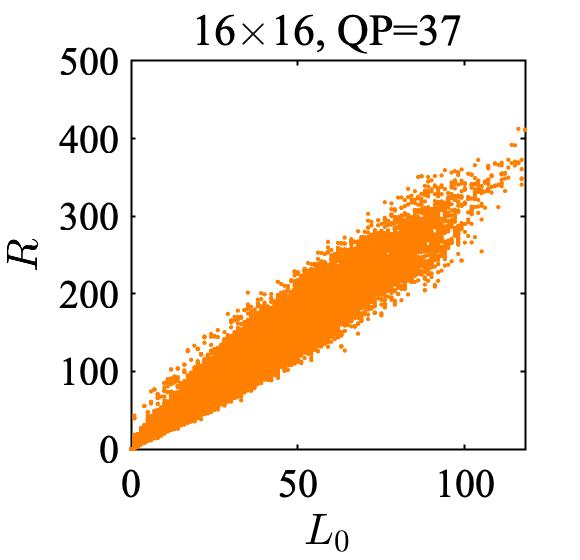}}\\
    \subfloat[]{\includegraphics[width=0.87in]{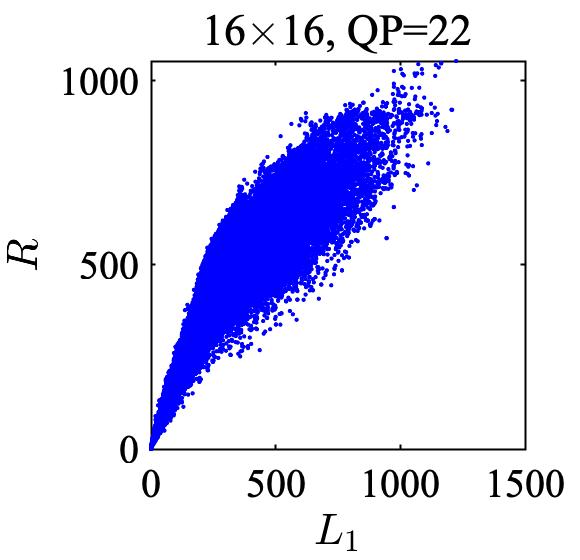}}
    \subfloat[]{\includegraphics[width=0.87in]{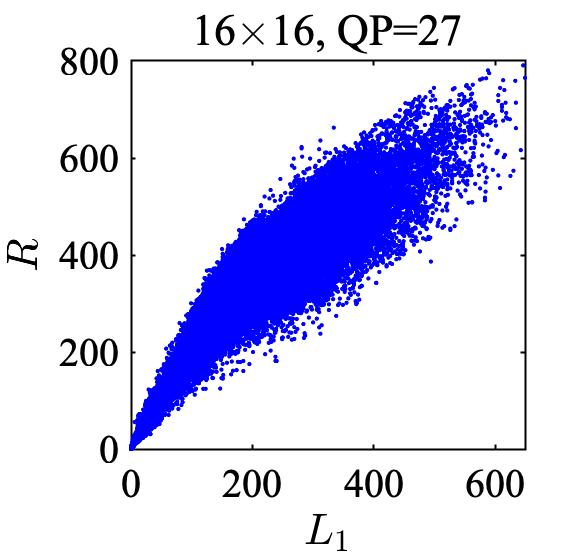}}
    \subfloat[]{\includegraphics[width=0.87in]{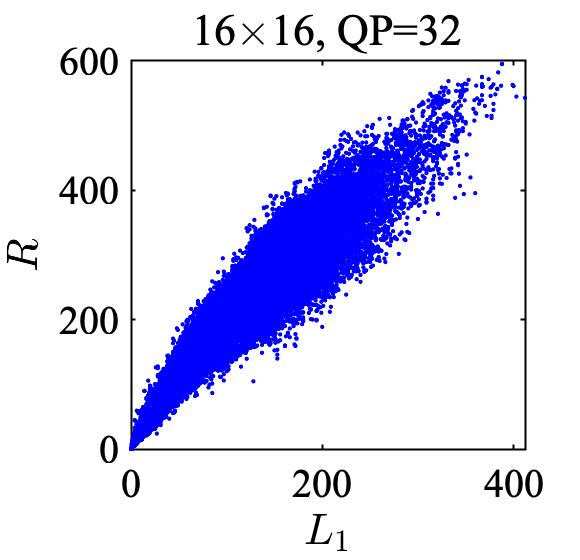}}
    \subfloat[]{\includegraphics[width=0.87in]{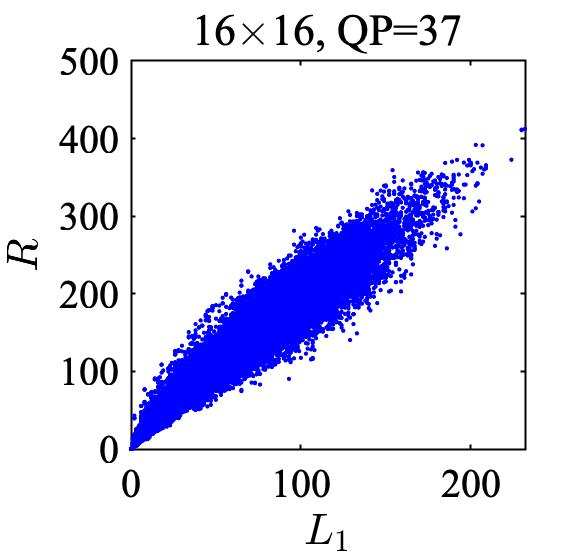}}\\
    \subfloat[]{\includegraphics[width=0.87in]{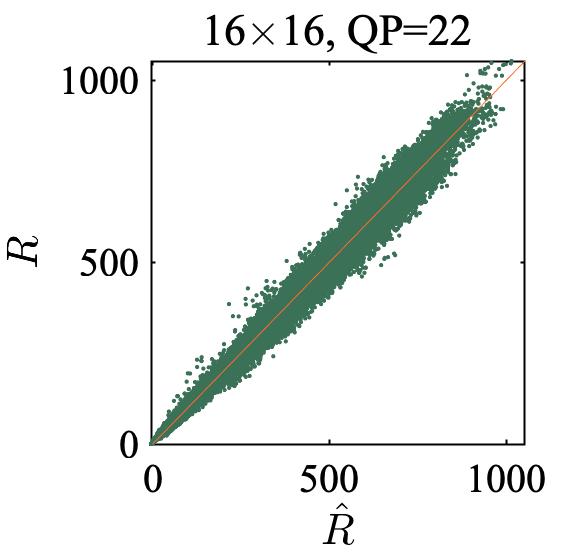}}
    \subfloat[]{\includegraphics[width=0.87in]{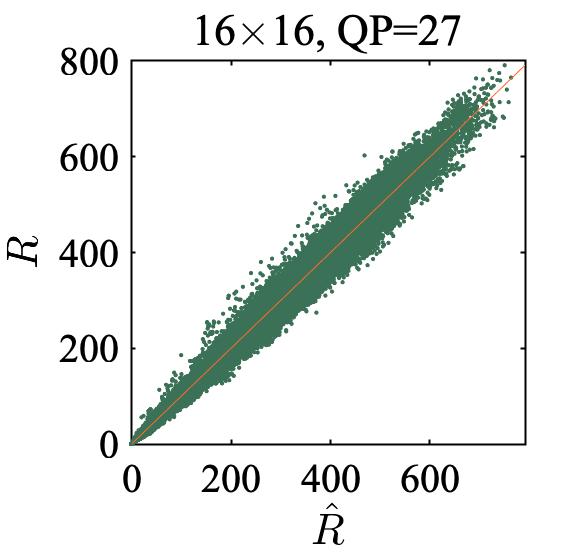}}
    \subfloat[]{\includegraphics[width=0.87in]{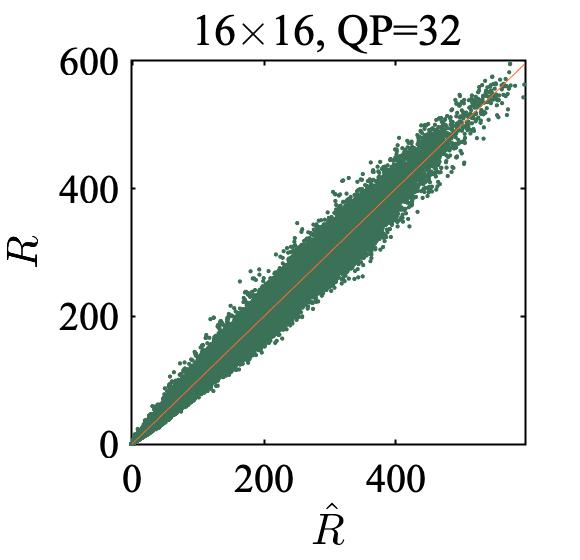}}
    \subfloat[]{\includegraphics[width=0.87in]{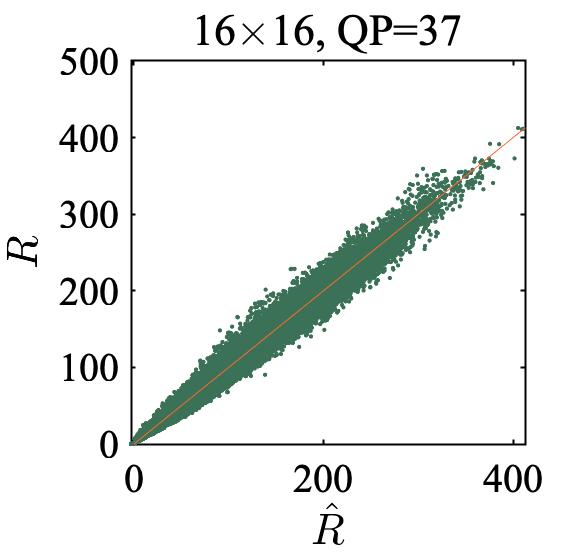}}
\caption{{Illustration of the actual coding bits $R$, $\ell_0$-norm, $\ell_1$-norm and estimated coding bits for the sequence ``RaceHorses''. (a-d) $R$ versus $\ell_0$-norm of quantized transform coefficients $L_0$; (e-h) $R$ versus $\ell_1$-norm of quantized transform coefficients $L_1$; (i-l) $R$ versus estimated coding bits $\hat{R}$.}}
\label{Fig_fit_l0_l1_racehorses}
\end{figure}

\subsection{Rate modeling}
The $\ell_0$-norm and $\ell_1$-norm of coefficients play complementary roles in approaching the number of coding bits, and merely employing $\ell_0$-norm or $\ell_1$-norm may lead to the biased approximation. First, the low bit rate assumption made by Eqn.~(\ref{taylorExp}) may not always hold. 
As illustrated in Fig.~\ref{Fig_Taylor}, 
when $P_{nz}$ is beyond 50\%, corresponding to high bit rate coding, the actual bits could be underestimated when adopting the $\ell_0$-norm only. 
In addition, it can be noticed that $\ell_0$-norm estimates the rate in a statistical manner without considering the individual level of coefficients. Obviously, larger coefficients should consume more coding bits, which cannot be well characterized by $\ell_0$-norm only. This also provides useful evidences regarding the incorporation of $\ell_1$-norm. On the other hand, $\ell_1$-norm only pays attention to the individual coding elements but ignores the dependencies and context in coefficients coding process. Therefore, it is possible to equip the rate model with both $\ell_0$-norm and $\ell_1$-norm. Moreover, the position information, especially the location of the last non-zero coefficients also influences the final coding bits. As such, the final rate model is given by,
\begin{align}
    \hat{R} = \alpha \cdot L_0 + \beta \cdot L_1 + \gamma \cdot R_{LP} + \epsilon,
    \label{R_L0_L1_LP}
\end{align}
where $\alpha$, $\beta$, $\gamma$ and $\epsilon$ denote the model parameters. In particular, the parameters $\alpha$ and $\beta$ which control the relationship between rate and \(\ell_0\)/\(\ell_1\)-norms of the quantized coefficients, highly rely on the $Q_{step}$ and block sizes. $L_0$ and $L_1$ represent the $\ell_0$-norm and $\ell_1$-norm of the current CU, respectively. In VVC, individual coordinate of the last significant coefficient is composed of a prefix and suffix, wherein the prefix is context coded with truncated unary bins and the suffix is bypass coded with fixed length bins. Here, the coding bit of the coordinates $(x, y)$ regarding the last non-zero coefficient is represented by $R_{LP}$, which is practically obtained by a look-up table. The relationship between the actual coding bits $R$ and the estimated coding bits $\hat{R}$ is illustrated in Fig.~\ref{Fig_fit_l0_l1_racehorses} with various QPs, showing that the model delivers high accuracy in modeling the rate in VVC.
\subsection{Distortion modeling}
To measure the quantization distortion, sum of square error (SSE) is adopted, and the SSE of a coding block can be straightforwardly represented as follows,
\begin{align}
    D &= \sum_{i = 0}^{W \times H - 1}\left[Q^{-1}\left(l_s^{(i)}\right) - C_s^{(i)}\right]^2 \nonumber\\
     &=\sum_{i = 0}^{W \times H - 1}\left[\left(Q_{step} \cdot l_s^{(i)}\right)^2 - 2\cdot Q_{step} \cdot l_s^{(i)}\cdot C_s^{(i)} + \left( C_s^{(i)}\right)^2\right],
    \label{dist_block}
\end{align}
where $Q^{-1}(\cdot)$ indicates the inverse quantization.

\section{Problem Formulation}

Scalar quantization has been widely used in video coding standards owing to its computational simplicity, as it generally employs one quantizer associated with a specific quantization step. 
TCQ was studied early in 1990~\cite{TCQ_1990}, which can be regarded as large-dimension vector quantization with constrained vector components and is capable of remedying the inevitable performance loss incurred by scalar quantization to some extent. The TCQ in VVC is implemented as dependent scalar quantization that simultaneously maintains two quantizers $Q_0$, $Q_1$ with four transition states. To be more specific, TCQ embeds quantization candidates in one block into trellis graph wherein the best quantization outcomes correspond to the path with the minimum RD cost. In this manner, the inter-dependencies of transform coefficients can be well exploited, and moreover, TCQ persuades the reconstruction vector to be more compact with augmented quantizers and candidates. As such, significantly better RD performance can be achieved.

\begin{figure}[t]
    \centering
    \includegraphics[width=5cm]{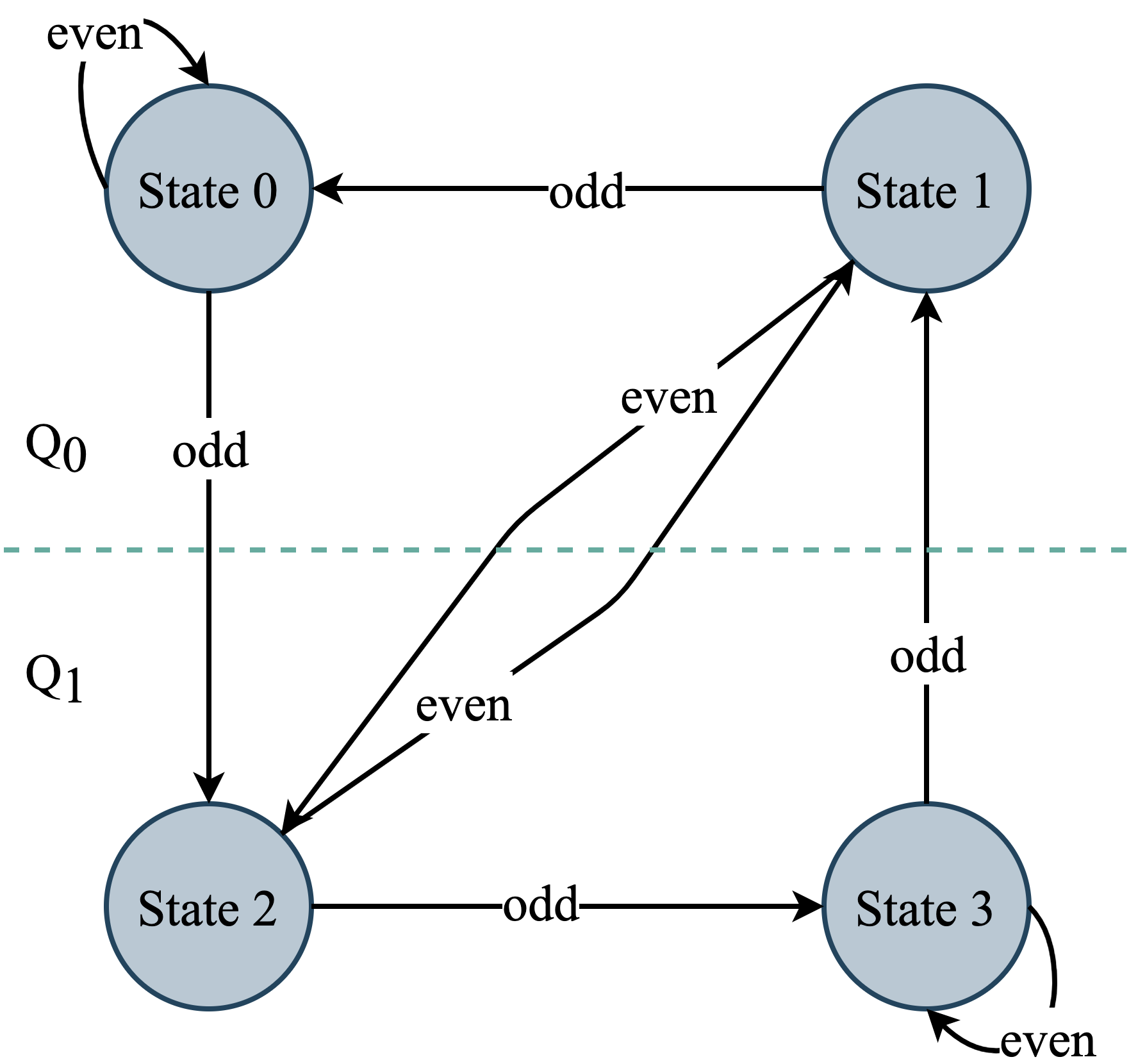}%
    \caption{Illustration of the state transition~\cite{vvc4}.}
     \label{Fig_State}
\end{figure}

More specifically, in TCQ of VVC, given a transform coefficient $C_s^{(i)}$, {several quantization candidates can be obtained based on the pre-quantization results.}
Subsequently, the quantization level $l^{(i)}$ is further converted into the quantization index $\tilde{l}^{(i)}$. Since the representation of quantization index is nearly half of the original quantization level, coding bits could be naturally saved. The parity of the current quantization index, as well as the current state, determine the state transition route and the quantizer for next coefficient, as illustrated in Fig.~\ref{Fig_State}.
Consequently, the reconstruction process of $Q_0$ is always associated to even times of quantization step $Q_{step}$, and $Q_1$ is bounded with odd times of $Q_{step}$. The reconstruction process is illustrated in Algorithm~\ref{rec}, {where $N_m$ denotes the number of quantization indices corresponding to the trellis stages within one coding block and $i$ represents the processing order}.

The quantization distortion and rate of individual coefficient are calculated and recorded for each trellis node following the scanning order. Given the current quantization index $\tilde{l}^{(i)}$ and transition state $St^{(i)}$, the quantization level can be reconstructed as follows,
\begin{align}
    l^{(i)} = 2\cdot \tilde{l}^{(i)} - (St^{(i)} >> 1)\cdot sign (\tilde{l}^{(i)}).
\end{align}
where $St^{(i)} >> 1$ typically specifies the utilized quantizer and in turn introduces even or odd multiples of quantization steps. As such, the distortion is given by~\cite{Trellis_DCC},
\begin{align}
    &D\left(\tilde{l}^{(i)}, St^{(i)}\right) = \nonumber \\
    &\left[C^{(i)} - Q_{step}\cdot (2\cdot \tilde{l}^{(i)} - (St^{(i)} >> 1)\cdot sign(\tilde{l}^{(i)}))\right]^2.
\end{align}

\begin{figure*}[t]
    \centering
    \includegraphics[width=16cm]{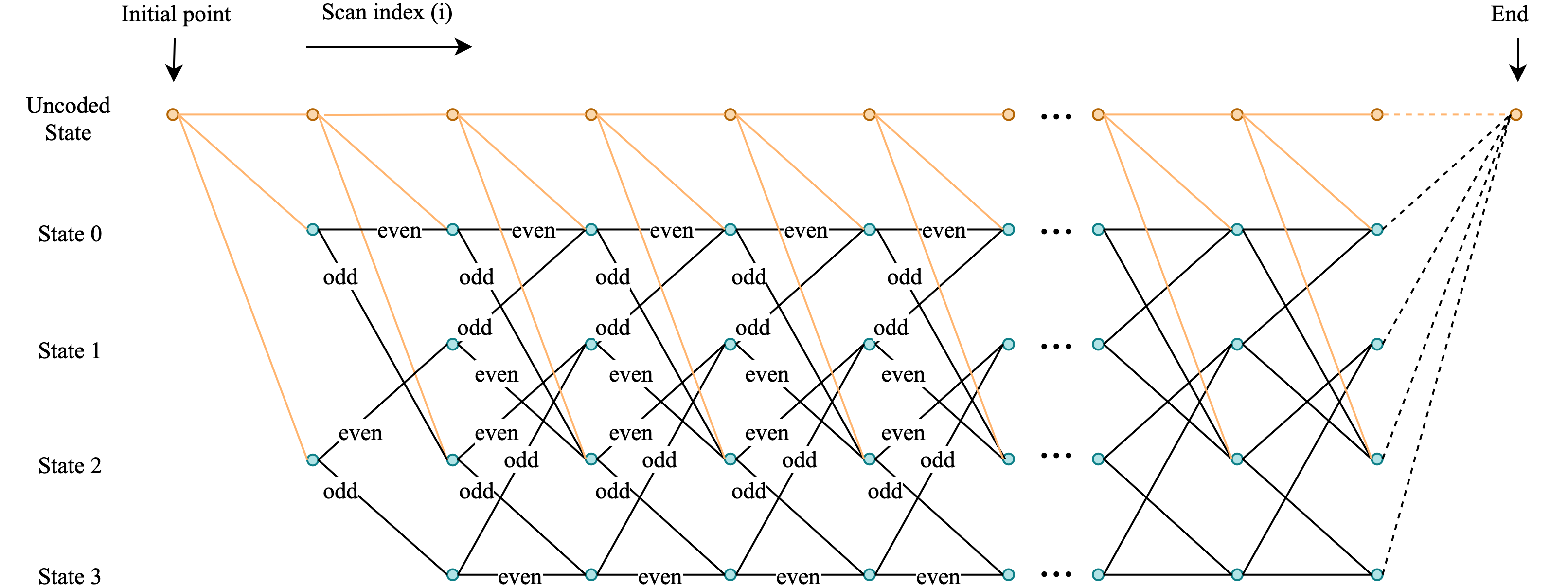}%
    \caption{Illustration of trellis graph~\cite{Trellis_DCC}.}
     \label{Fig_Trellis}
\end{figure*}

The absolute of quantization index $|\tilde{l}^{(i)}|$ is entropy coded by signaling the syntax $sig$, $gt1$, $par$ and $gtx$ with regular mode. Also, the remaining levels denoted by $rem$ are binarized with Golomb-Rice code and coded in bypass mode. 
Despite of the aforementioned four states that are involved in the state transition loop, a special state termed as ``uncoded'' state is introduced, which attempts to truncate the residuals locating in the high frequency domain, in an effort to further save the coding bits. 
An exemplified trellis graph of one coding block is illustrated in Fig.~\ref{Fig_Trellis}. The switching from ``uncoded'' state to State 0 or State 2 is only allowed when encountering non-zero quantization indices.

Following the scanning order, the cost of each individual stage is accumulated along the transition path until attaining the end of the block. Typically, there are multiple enter-paths with different accumulative costs attaining to the same node. Only one path with the lowest cost is retained as the survivor path. Considering the reverse scanning order, the cost accumulating can be described as follows,
\begin{align}
    J^{(i)} = J^{(i+1)} + D(\tilde{l}^{(i)}, St^{(i)}) + \lambda \cdot R(\tilde{l}^{(i)}).
\end{align}
In particular, if the state transition is from one ``uncoded'' state to another ``uncoded'' state, the RD cost is iterated as follows,
\begin{align}
    J^{(i)} = J^{(i+1)} + D(0, 0).
\end{align}
By contrast, if the state switches from ``uncoded'' state to State 0 or State 2, $J^{(i)}$ can be calculated as follows,
\begin{align}
    J^{(i)} &= J^{(i+1)} + D(\tilde{l}^{(i)}, St^{(i)}) \nonumber \\
    &+ \lambda \cdot [R(\tilde{l}^{(i)}) + R_{cbf}, + R_{LP}(x^{(i)}, y^{(i)})],
\end{align}
where $R_{cbf}$ denotes the bits used for representing the variation of the coded block flag ($cbf$, from 0 to 1). $R_{LP}$ denotes the bits regarding to the position of the last (first traversed) non-zero coefficient, the coordinators of which are represented with $x^{(i)}$ and $y^{(i)}$. 

Such cost accumulation, path comparison and the optimal branch selection process can be regarded as the add-compare-select (ACS)~\cite{HaibinYin}. Moreover, RD cost calculation is conducted in branch metric unit (BMU). Essentially, the goal of TCQ is to find the optimal quantization solution that can achieve the minimum RD cost for the whole coding block. Viterbi algorithm is used for the optimal path searching. Multiple routes are available at each stage, where each route represents the state transition invited by the quantization index $\tilde{l}^{(i)}$ and individual transform coefficient can be regarded as the stage of trellis graph.  After attempting all the paths linked with one node, only one path with the lowest accumulated RD cost that connects to the destination node will be retained.

The complexity of TCQ is mainly attributed to three factors. 
The first one is the total number of trellis stages $N_{m}$, which generally corresponds to the number of coefficients within one block. 
The second one is the number of branches $N_b^{(i)}$ linked to a node, which attributes to the quantity of quantization candidates. The third one is the number of states at each stage $N_{state}^{(i)}$. Since the quantization indices are grouped based on the parity, the number of branches linked to an individual node is halved compared to the full connected trellis, and candidate ``0'' should be additionally counted. Supposing $N_l^{(i)}$ quantization candidates are available for a single transform coefficient, the branch complexity $N_b^{(i)}$ can be described as,
\begin{align}
    N_b^{(i)} = N_l^{(i)} / 2 + 1.
\end{align}
For a typical coding block with $N_m$ coefficients, the total branch count of TCQ can be formulated as,
\begin{align}
   N_{TCQ} &=  \sum_{i = 1}^{N_{m}} N_b^{(i)} \times N_{state}^{(i)}.
   \label{NTCQ}
\end{align}
Consequently, the complexity is proportional to the stage number $N_m$ and branch number $N_b^{(i)}$. We demonstrate the theoretical computational complexity in Table~\ref{complex_1}. To better handle those three problems while maintaining the efficiency of TCQ, we apply the established rate and distortion models to achieve low complexity TCQ for VVC.

\begin{table*}[t]
  \centering
  \scriptsize
  \caption{Complexity analyses of TCQ}
    \begin{tabular}{cccc|ccc}
    \toprule
    \multirow{2}[4]{*}{Module} & \multirow{2}[4]{*}{Branch} & \multicolumn{2}{c|}{BMU} & \multicolumn{3}{c}{ACS} \\
\cmidrule{3-7}          &       & Distortion & Rate  & Add   & Compare & Select \\
    \midrule
    TCQ   & $N_{TCQ}$  &   $\sum_{i = 1}^{N_{m}} N_l^{(i)}$    &  $\sum_{i = 1}^{N_{m}} N_l^{(i)}$     &   $N_{TCQ}$    &   $\sum_{i=1}^{N_{m}}(N_b^{(i)}-1)\cdot N_{state}^{(i)}$    & $\sum_{i=1}^{N_{m}}(N_b^{(i)}-1)\cdot N_{state}^{(i)}$ \\
    \bottomrule
    \end{tabular}%
  \label{complex_1}%
\end{table*}%

\begin{algorithm}[t]
\caption{Reconstruction of transform coefficients with trellis coded quantization~\cite{vvc4}}
\label{rec}
    \begin{algorithmic}[1]
        \Require 
            $Q_{step}$, $\tilde{l}^{(*)}$;
        \Ensure 
            Dequantization results: $\hat{C}^{(*)}$
        \State Initialize state: $St^{(N_m)} \gets 0$;
        \For {$i = N_m$; $i >= 0$; $i--$}
            \State $\hat{C}^{(i)} = (2\cdot \tilde{l}^{(i)} - (St^{(i)} >> 1) \cdot sign(\tilde{l}^{(i)})) \cdot Q_{step}$;
            \State $St^{(i-1)} = ( 32040 >> ((St^{(i)}<<2)+((\tilde{l}^{(i)} \& 1)<<1)) ) \& 3;$
        \EndFor
    \end{algorithmic}
\end{algorithm}

\begin{figure*}[!htb]
    \centering
    \subfloat[]{\includegraphics[width=1.5in]{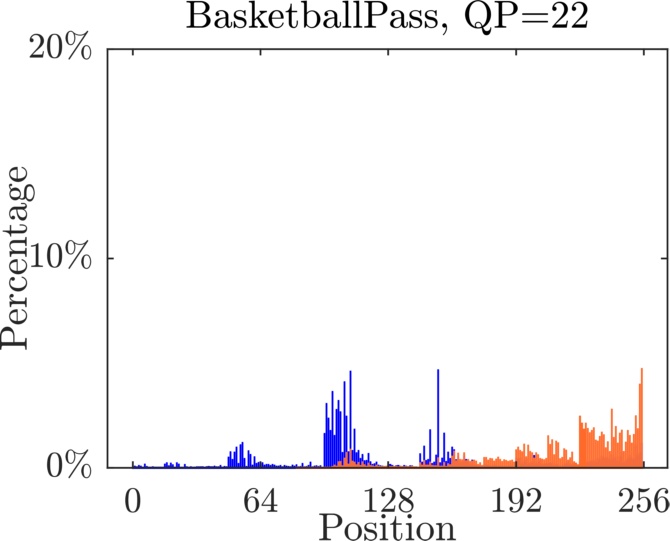}}%BasketballPass;
    \hfil
    \subfloat[]{\includegraphics[width=1.5in]{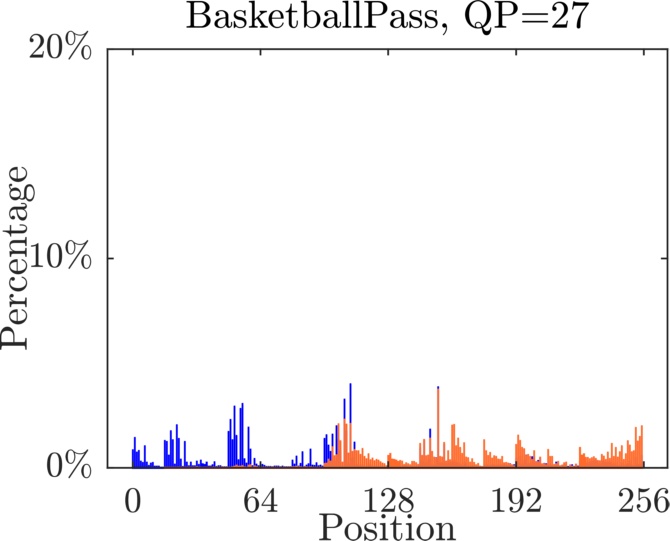}}%BasketballPass;
    \hfil
    \subfloat[]{\includegraphics[width=1.5in]{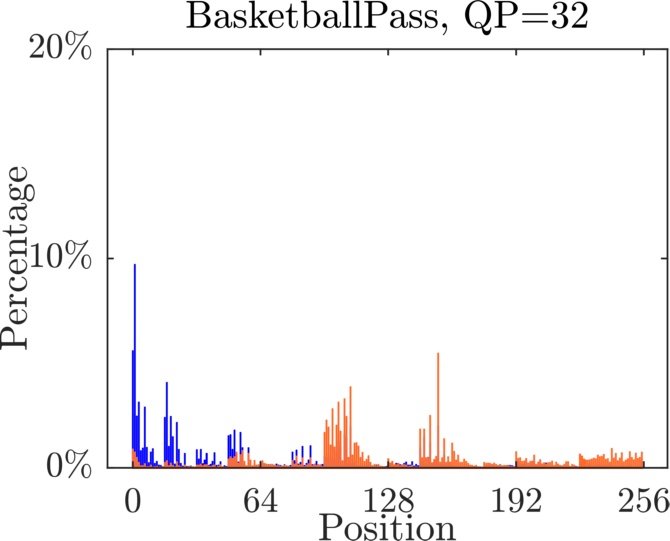}}%BasketballPass;
    \hfil
     \subfloat[]{\includegraphics[width=1.5in]{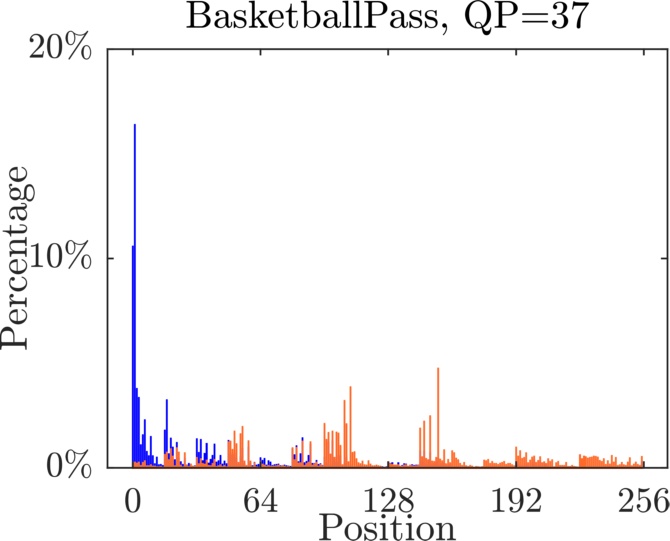}}\\%BasketballPass;
\caption{{Illustration of the position distribution of the last non-zero coefficient index in 16 $\times$16 coding blocks under AI configuration of VVC. Both HDQ and TCQ are considered, where HDQ corresponds to orange bars and TCQ is represented with blue bars.}}
\label{lastPos_static_AI}
\end{figure*}

\section{Low Complexity Trellis-Coded Quantization}
\subsection{Determination of the Trellis Departure Point}
{We first propose to softly decide the trellis departure point based on the rate and distortion models, with the goal of shrinking the total number of trellis stages $N_m$.}
Quantization and residual coefficient coding are conducted based on coefficient group (CG) with reverse scanning order. As such, coefficients from the bottom right to the left-top within one coding block are orderly mapped into trellis along with accessible states. 
The starting point, which serves as the first non-zero point during traversing of the trellis graph, plays critical roles in TCQ. 
It is widely acknowledged that the coefficients locating within the high frequency regions tend to be quantized to zeros in a soft way in the sense of rate-distortion optimization. To manifest this, statistical experiments are conducted to exploit the distribution of the last non-zero coefficient position with the HDQ and TCQ in $16 \times 16$ coding blocks, as illustrated in Fig.~\ref{lastPos_static_AI}. It can be observed that the positions of the last non-zero coefficient are
concentrated on smaller scan indices with TCQ compared to HDQ. With the increase of QP, such distribution differences become more apparent.

As such, by combining the rate-distortion models and TCQ, we propose an algorithm that allows us to accurately determine the trellis departure point in an elegant and low cost way. More specifically, we cast this problem into the comparisons of the RD cost, as the RD cost differences associated with two contiguous non-zero quantized coefficients can be measured and compared to determine the optimal trellis starting point. With the proposed algorithm, the initial point of the trellis graph can be postponed, leading to the shrinkage of the total stages as well as computational complexity in determining the optimal quantized coefficients of the given coding block.

Supposing $i$ and $j$ are two typical positions which satisfy the following constraints,
\begin{align}
    &i, j \in [0, W\times H - 1], \quad  
    i > j,\quad 
    l^{(i)} \neq 0, \quad l^{(j)} \neq 0.  \quad\nonumber \\
    &l^{(k)} = 0, \quad \text{if }k \in [j+1, i-1].
    \label{i_j}
\end{align}
{Here, we use $C^{(*)}$ and $l^{(*)}$ to represent the absolute value of $C_s^{(*)}$ and $l_s^{(*)}$, respectively.}
Since coefficient $i$ is ahead of $j$ during traversing, the quantization result $l^{(i)}$ should be zero if position $j$ serves as the initial point of trellis. 

{The rate estimation model in Eqn.~(\ref{R_L0_L1_LP}) is employed where the total rates of a block can be represented with $R^{(0, j)}$ and $R^{(0, i)}$ when position $j$ and $i$ serve as the trellis initial point, respectively.
As such, the rate difference is formulated as follows,}
\begin{align}
    \Delta R^{(j, i)} &= R^{(0, j)} - R^{(0, i)} \nonumber \\
    & = - [\alpha + \beta \cdot \tilde{l}^{(i)} + \gamma \cdot (R_{LP}^{(i)} - R_{LP}^{(j)}) ].
    \label{deltaR}
\end{align}
where $\tilde{l}^{(i)}$ denotes the quantization index of ${l}^{(i)}$.

Regarding the distortion, we further formulate $\Delta D^{(j, i)}$ with $D^{(0, j)}$ and $D^{(0, i)}$, where $D^{(0, j)}$ denotes the overall distortion when the coefficient at position $j$ is regarded as the trellis starting point. The derivations of $D^{(0, j)}$ and $D^{(0, i)}$ are given by,
\begin{align}
    D^{(0, j)} &= \sum_{k=0}^{j} [Q^{-1}(l^{(k)}) - C^{(k)}]^2  +\sum_{k=j+1}^{W\times H - 1} (C^{(k)})^2.
\end{align}
\begin{align}
    D^{(0, i)} &= \sum_{k=0}^{j} [Q^{-1}(l^{(k)}) - C^{(k)}]^2  +\sum_{k=j+1}^{i-1} (C^{(k)})^2  \nonumber \\
    &+ [Q^{-1}(l^{(i)}) - C^{(i)}]^2 + \sum_{k=i+1}^{W\times H - 1} (C^{(k)})^2.
\end{align}
As such, $\Delta D^{(j, i)}$ can be expressed as,
\begin{align}
    \Delta D^{(j, i)} &= D^{(0, j)} - D^{(0, i)} \nonumber \\
    &= -[(Q_{step} \cdot l^{(i)})^2 - 2\cdot Q_{step} \cdot l^{(i)}\cdot C^{(i)}].
    \label{dist_twopoint_simp}
\end{align}

The RD cost difference $\Delta J^{(j, i)}$ that characterizes the variation of RD cost by {removing $i$ out of the trellis} can be derived as follows,
\begin{equation}
    \Delta J^{(j, i)} = \Delta D^{(j, i)} + \lambda \cdot \Delta R^{(j, i)}.
    \label{deltaJ}
\end{equation}
The non-zero coefficient at position $i$ can be eliminated from the trellis according to $\Delta J^{(j, i)}$. In other words, if $\Delta J^{(j, i)}\leq0$, it is unnecessary to involve the coefficient at position $i$ in the trellis. 

As such, by combining Eqn.~(\ref{deltaR}), Eqn.~(\ref{dist_twopoint_simp}) and Eqn.~(\ref{deltaJ}), an RD-based threshold with respect to $C^{(i)}$ can be derived as follows,
\begin{equation}
C^{(i)} \leq T,
\label{thd}
\end{equation}
where
\begin{align}
 T &=
\frac{1}{2} \cdot \left(Q_{step} \cdot l^{(i)} + \frac{\lambda \cdot (\alpha + \beta \cdot \tilde{l}^{(i)} + \gamma \cdot (R_{LP}^{(i)} - R_{LP}^{(j)}))}{Q_{step} \cdot l^{(i)}}\right).
\end{align}
Assuming that the rate for representing the last position $i$ is larger than or equal to that of position $j$, by approximating $\tilde{l}^{(i)}$ with $0.5l^{(i)}$ , $R_{LP}^{(i)}$ with $R_{LP}^{(j)}$ , and substituting 
$\lambda$ with $\phi \cdot Q_{step}^2$, $T$ can be simplified as follows,
\begin{align}
    T \approx \frac{1}{2} \cdot  Q_{step} \cdot \left(l^{(i)} + \frac{\phi \cdot \alpha}{l^{(i)}} + \frac{\phi}{2} \cdot \beta \right),
\end{align}
where $\phi$ is a multiplication factor that can be obtained according to the VVC configuration. The theoretical minimum value of $T$ can be obtained as follows,
\begin{align}
    T \geq Q_{step} \cdot \mathcal{K},
\end{align}
where 
\begin{align}
    \mathcal{K} = \sqrt{\phi \cdot \alpha } + \frac{\phi}{4} \cdot \beta.
    \label{K_factor}
\end{align}
Herein, the minimum value of $T$ is adopted as the threshold.
If the condition of Eqn. (\ref{thd}) is satisfied, the associated quantization coefficient can be directly determined as zero and further removed from the trellis graph. In this way, the trellis starting point can be postponed to the next non-zero coefficient. Otherwise, $C^{(i)}$ is regarded as the starting point of the trellis graph.

\begin{figure}[t]
    \centering
    \includegraphics[width=8cm]{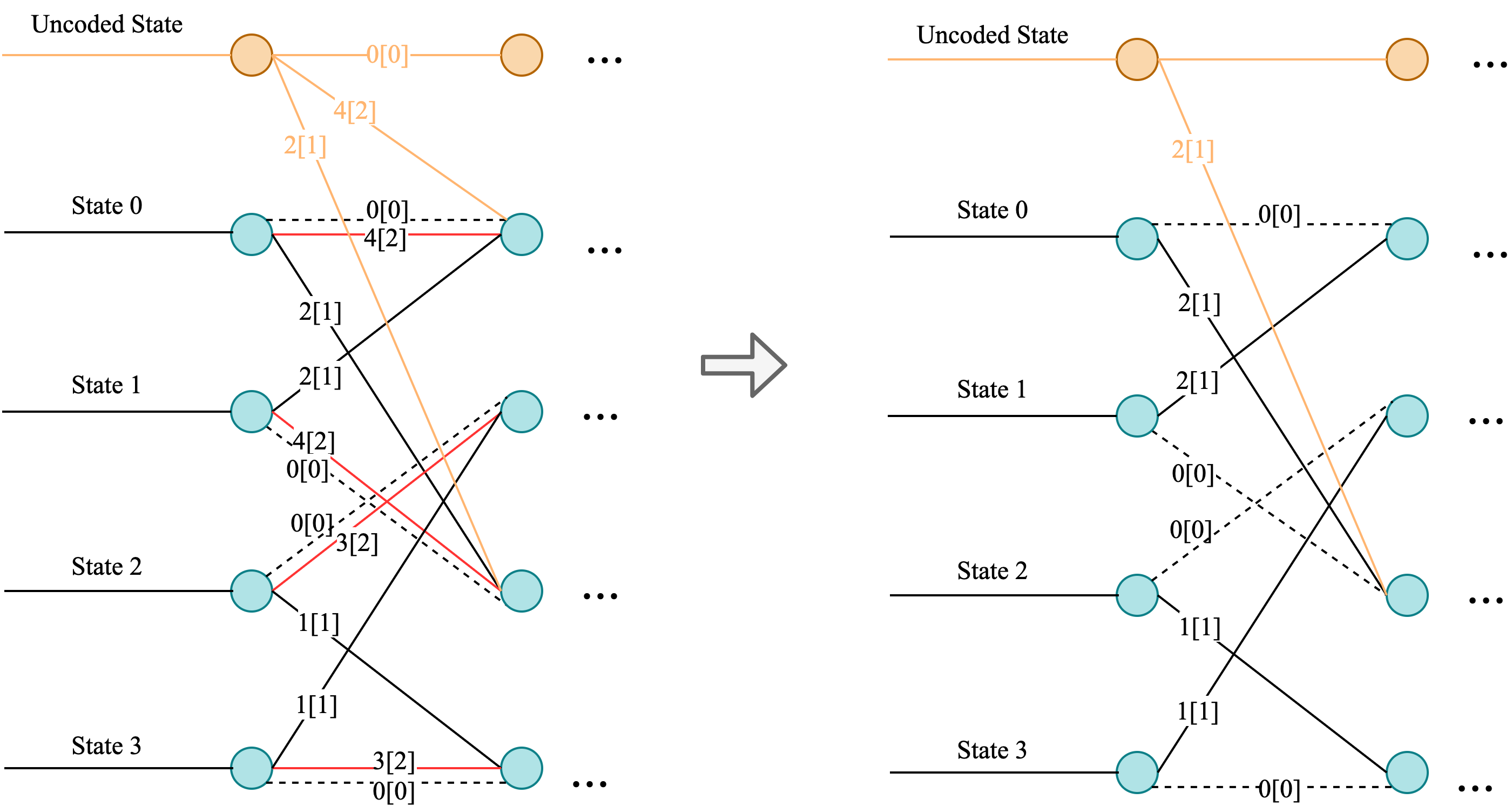}%
    \caption{Illustration of the trellis pruning for the larger quantization candidates.}
     \label{Fig_pruning_1}
\end{figure}

\begin{figure}[t]
    \centering
    \includegraphics[width=8cm]{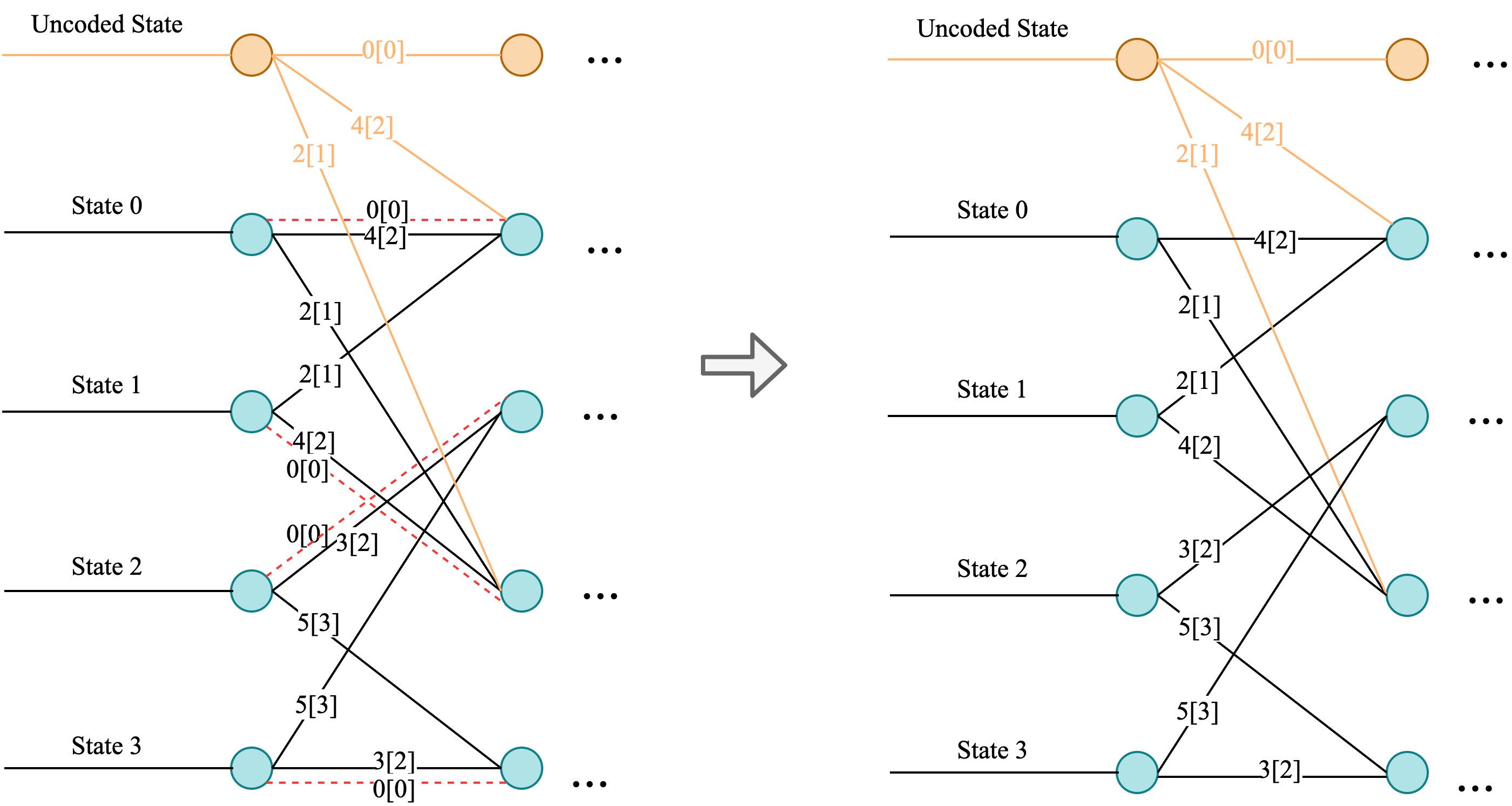}%
    \caption{Illustration of the trellis pruning for the smaller quantization candidates.}
     \label{Fig_pruning_2}
\end{figure}

% Table generated by Excel2LaTeX from sheet 'adaptive'
\begin{table*}[t]
  \centering
  \scriptsize
  \caption{Performance of the proposed scheme by postponing the trellis departure point under AI and RA configurations}
    \begin{tabular}{cc|ccccc|ccccc}
    \toprule
    \multicolumn{1}{c|}{\multirow{2}[2]{*}{Class}} & \multirow{2}[2]{*}{Sequence} & \multicolumn{5}{c|}{AI}               & \multicolumn{5}{c}{RA} \\
    \multicolumn{1}{c|}{} &       & BD-Rate(Y)     & BD-Rate(U)     & BD-Rate(V)     & $TS_Q$     & $TS_{Enc}$     & BD-Rate(Y)     & BD-Rate(U)     & BD-Rate(V)     & $TS_Q$     & $TS_{Enc}$ \\
    \midrule
    \multicolumn{1}{c|}{\multirow{3}[2]{*}{A1}} & Tango2 & 0.07\% & 0.23\% & 0.32\% & 34\%  & 13\%  & 0.01\% & 0.20\% & 0.26\% & 32\%  & 3\% \\
    \multicolumn{1}{c|}{} & FoodMarket4 & 0.09\% & 0.22\% & 0.16\% & 26\%  & 10\%  & 0.04\% & -0.14\% & -0.08\% & 19\%  & 0\% \\
    \multicolumn{1}{c|}{} & Campfire & 0.05\% & 0.09\% & 0.02\% & 24\%  & 10\%  & 0.01\% & 0.14\% & 0.10\% & 23\%  & 5\% \\
    \midrule
    \multicolumn{1}{c|}{\multirow{3}[2]{*}{A2}} & CatRobot1 & 0.11\% & 0.18\% & 0.14\% & 28\%  & 11\%  & -0.02\% & 0.01\% & 0.13\% & 32\%  & 2\% \\
    \multicolumn{1}{c|}{} & DaylightRoad2 & 0.16\% & 0.41\% & 0.20\% & 26\%  & 11\%  & 0.07\% & -0.14\% & -0.09\% & 33\%  & 1\% \\
    \multicolumn{1}{c|}{} & ParkRunning3 & 0.03\% & 0.13\% & 0.17\% & 17\%  & 7\%   & 0.01\% & 0.12\% & 0.16\% & 21\%  & 1\% \\
    \midrule
    \multicolumn{1}{c|}{\multirow{5}[2]{*}{B}} & MarketPlace & 0.03\% & 0.15\% & 0.38\% & 22\%  & 10\%  & 0.04\% & 0.23\% & 0.41\% & 26\%  & 5\% \\
    \multicolumn{1}{c|}{} & RitualDance & 0.06\% & 0.28\% & 0.15\% & 21\%  & 9\%   & 0.04\% & -0.22\% & -0.18\% & 22\%  & 5\% \\
    \multicolumn{1}{c|}{} & Cactus & 0.07\% & 0.09\% & 0.21\% & 22\%  & 10\%  & 0.03\% & 0.10\% & 0.29\% & 27\%  & 6\% \\
    \multicolumn{1}{c|}{} & BasketballDrive & 0.10\% & 0.10\% & 0.16\% & 27\%  & 12\%  & 0.05\% & -0.46\% & -0.01\% & 27\%  & 5\% \\
    \multicolumn{1}{c|}{} & BQTerrace & 0.06\% & 0.10\% & 0.23\% & 17\%  & 8\%   & 0.07\% & 0.79\% & 0.16\% & 23\%  & 4\% \\
    \midrule
    \multicolumn{1}{c|}{\multirow{4}[2]{*}{C}} & BasketballDrill & 0.14\% & 0.05\% & 0.14\% & 23\%  & 11\%  & 0.05\% & -0.05\% & 0.07\% & 27\%  & 7\% \\
    \multicolumn{1}{c|}{} & BQMall & 0.08\% & 0.10\% & 0.15\% & 20\%  & 10\%  & 0.05\% & -0.35\% & -0.25\% & 23\%  & 5\% \\
    \multicolumn{1}{c|}{} & PartyScene & 0.09\% & -0.03\% & 0.15\% & 13\%  & 6\%   & 0.08\% & 0.10\% & 0.30\% & 19\%  & 6\% \\
    \multicolumn{1}{c|}{} & RaceHorses & 0.08\% & 0.05\% & 0.00\% & 17\%  & 8\%   & 0.03\% & 0.19\% & 0.21\% & 21\%  & 5\% \\
    \midrule
    \multicolumn{1}{c|}{\multirow{4}[2]{*}{D}} & BasketballPass & 0.07\% & 0.09\% & 0.26\% & 18\%  & 8\%   & 0.08\% & -0.30\% & -0.07\% & 19\%  & 4\% \\
    \multicolumn{1}{c|}{} & BQSquare & 0.10\% & -0.02\% & -0.20\% & 12\%  & 5\%   & 0.00\% & -0.53\% & -1.09\% & 17\%  & 3\% \\
    \multicolumn{1}{c|}{} & BlowingBubbles & 0.11\% & -0.22\% & 0.37\% & 13\%  & 7\%   & 0.17\% & 0.69\% & 0.41\% & 20\%  & 4\% \\
    \multicolumn{1}{c|}{} & RaceHorses & 0.14\% & -0.25\% & -0.67\% & 15\%  & 6\%   & -0.05\% & -0.28\% & 0.55\% & 19\%  & 3\% \\
    \midrule
    \multicolumn{1}{c|}{\multirow{3}[2]{*}{E}} & FourPeople & 0.11\% & 0.17\% & 0.20\% & 19\%  & 9\%   & -     & -     & -     & -     & - \\
    \multicolumn{1}{c|}{} & Johnny & 0.09\% & 0.40\% & 0.11\% & 21\%  & 9\%   & -     & -     & -     & -     & - \\
    \multicolumn{1}{c|}{} & KristenAndSara & 0.08\% & 0.05\% & 0.13\% & 20\%  & 8\%   & -     & -     & -     & -     & - \\
    \midrule
    \multicolumn{1}{c|}{\multirow{4}[2]{*}{F}} & BasketballDrillText & 0.11\% & 0.07\% & 0.07\% & 19\%  & 8\%   & 0.05\% & 0.09\% & 0.10\% & 25\%  & 6\% \\
    \multicolumn{1}{c|}{} & ArenaOfValor & 0.11\% & 0.06\% & 0.18\% & 17\%  & 4\%   & 0.10\% & 0.09\% & 0.18\% & 22\%  & 6\% \\
    \multicolumn{1}{c|}{} & SlideEditing & 0.01\% & 0.09\% & 0.21\% & 7\%   & 4\%   & -0.03\% & 0.07\% & -0.02\% & 10\%  & 2\% \\
    \multicolumn{1}{c|}{} & SlideShow & 0.09\% & 0.50\% & 0.35\% & 12\%  & 4\%   & 0.18\% & -0.19\% & -0.12\% & 12\%  & 2\% \\
    \midrule
    \multicolumn{2}{c|}{Class A1} & 0.07\% & 0.18\% & 0.16\% & 28\%  & 11\%  & 0.02\% & 0.06\% & 0.09\% & 25\%  & 3\% \\
    \multicolumn{2}{c|}{Class A2} & 0.10\% & 0.24\% & 0.17\% & 23\%  & 10\%  & 0.02\% & 0.00\% & 0.07\% & 29\%  & 1\% \\
    \multicolumn{2}{c|}{Class B} & 0.07\% & 0.14\% & 0.23\% & 22\%  & 10\%  & 0.04\% & 0.09\% & 0.13\% & 25\%  & 5\% \\
    \multicolumn{2}{c|}{Class C} & 0.10\% & 0.04\% & 0.11\% & 18\%  & 9\%   & 0.05\% & -0.03\% & 0.08\% & 23\%  & 5\% \\
    \multicolumn{2}{c|}{Class E} & 0.09\% & 0.21\% & 0.15\% & 20\%  & 9\%   & -     & -     & -     & -     & - \\
    \midrule
    \multicolumn{2}{c|}{\textbf{Overall}} & \textbf{0.09\%} & \textbf{0.15\%} & \textbf{0.17\%} & \textbf{22\%} & \textbf{10\%} & \textbf{0.03\%} & \textbf{0.03\%} & \textbf{0.08\%} & \textbf{25\%} & \textbf{4\%} \\
    \midrule
    \multicolumn{2}{c|}{Class D} & 0.10\% & -0.10\% & -0.06\% & 14\%  & 7\%   & 0.05\% & -0.11\% & -0.05\% & 19\%  & 3\% \\
    \multicolumn{2}{c|}{Class F} & 0.08\% & 0.18\% & 0.20\% & 14\%  & 5\%   & 0.07\% & 0.02\% & 0.04\% & 17\%  & 4\% \\
    \bottomrule
    \end{tabular}%
  \label{TAB_ADAPTIVE}%
\end{table*}%
% Table generated by Excel2LaTeX from sheet 'Prune'
\begin{table*}[htbp]
  \centering
  \scriptsize
  \caption{Performance of the proposed trellis pruning method under AI and RA configurations}
    \begin{tabular}{cc|ccccc|ccccc}
    \toprule
    \multicolumn{1}{c|}{\multirow{2}[2]{*}{Class}} & \multirow{2}[2]{*}{Sequence} & \multicolumn{5}{c|}{AI}               & \multicolumn{5}{c}{RA} \\
    \multicolumn{1}{c|}{} &       & BD-Rate(Y)     & BD-Rate(U)     & BD-Rate(V)     & $TS_Q$    & $TS_{Enc}$    & BD-Rate(Y)     & BD-Rate(U)     & BD-Rate(V)     & $TS_Q$     & $TS_{Enc}$ \\
    \midrule
    \multicolumn{1}{c|}{\multirow{3}[2]{*}{A1}} & Tango2 & 0.04\% & 0.03\% & 0.06\% & 3\%   & 1\%   & 0.01\% & -0.05\% & 0.27\% & 5\%   & 4\% \\
    \multicolumn{1}{c|}{} & FoodMarket4 & 0.06\% & 0.03\% & 0.11\% & 2\%   & 1\%   & 0.00\% & -0.20\% & 0.09\% & 3\%   & 4\% \\
    \multicolumn{1}{c|}{} & Campfire & 0.01\% & 0.05\% & 0.01\% & 3\%   & 1\%   & 0.00\% & 0.05\% & -0.09\% & 4\%   & 1\% \\
    \midrule
    \multicolumn{1}{c|}{\multirow{3}[2]{*}{A2}} & CatRobot1 & 0.04\% & 0.13\% & -0.06\% & 4\%   & 1\%   & -0.01\% & -0.12\% & 0.00\% & 4\%   & 3\% \\
    \multicolumn{1}{c|}{} & DaylightRoad2 & 0.02\% & 0.21\% & 0.08\% & 4\%   & 2\%   & 0.01\% & 0.01\% & -0.04\% & 7\%   & 3\% \\
    \multicolumn{1}{c|}{} & ParkRunning3 & 0.05\% & 0.05\% & 0.02\% & 3\%   & 1\%   & 0.01\% & 0.04\% & 0.02\% & 3\%   & 7\% \\
    \midrule
    \multicolumn{1}{c|}{\multirow{5}[2]{*}{B}} & MarketPlace & 0.03\% & 0.07\% & 0.19\% & 2\%   & 1\%   & -0.01\% & 0.00\% & -0.22\% & 2\%   & 2\% \\
    \multicolumn{1}{c|}{} & RitualDance & 0.02\% & 0.30\% & -0.01\% & 3\%   & 2\%   & 0.01\% & 0.03\% & 0.00\% & 2\%   & 2\% \\
    \multicolumn{1}{c|}{} & Cactus & 0.03\% & -0.01\% & 0.11\% & 3\%   & 1\%   & 0.03\% & -0.23\% & -0.03\% & 3\%   & 0\% \\
    \multicolumn{1}{c|}{} & BasketballDrive & 0.04\% & 0.09\% & -0.07\% & 4\%   & 1\%   & 0.01\% & -0.48\% & -0.15\% & 3\%   & 0\% \\
    \multicolumn{1}{c|}{} & BQTerrace & 0.04\% & -0.15\% & 0.12\% & 2\%   & 1\%   & 0.05\% & 0.57\% & -0.53\% & 3\%   & 0\% \\
    \midrule
    \multicolumn{1}{c|}{\multirow{4}[2]{*}{C}} & BasketballDrill & 0.03\% & -0.06\% & 0.09\% & 0\%   & 0\%   & 0.04\% & 0.07\% & -0.17\% & 2\%   & 0\% \\
    \multicolumn{1}{c|}{} & BQMall & 0.01\% & -0.04\% & -0.19\% & 3\%   & 0\%   & 0.03\% & -0.13\% & -0.26\% & 1\%   & -1\% \\
    \multicolumn{1}{c|}{} & PartyScene & 0.03\% & -0.01\% & -0.03\% & 0\%   & 2\%   & 0.02\% & 0.15\% & -0.28\% & 0\%   & -1\% \\
    \multicolumn{1}{c|}{} & RaceHorses & 0.06\% & -0.09\% & 0.02\% & 2\%   & 0\%   & 0.02\% & -0.21\% & 0.00\% & -1\%  & -4\% \\
    \midrule
    \multicolumn{1}{c|}{\multirow{4}[2]{*}{D}} & BasketballPass & 0.03\% & 0.04\% & 0.10\% & 1\%   & 1\%   & -0.06\% & -1.01\% & 0.02\% & 0\%   & 0\% \\
    \multicolumn{1}{c|}{} & BQSquare & 0.06\% & -0.21\% & -0.10\% & 1\%   & 0\%   & -0.05\% & -0.87\% & -1.24\% & 0\%   & 0\% \\
    \multicolumn{1}{c|}{} & BlowingBubbles & 0.06\% & -0.11\% & -0.17\% & 0\%   & 1\%   & 0.09\% & 0.18\% & 0.13\% & 0\%   & 0\% \\
    \multicolumn{1}{c|}{} & RaceHorses & 0.04\% & 0.02\% & -0.44\% & 1\%   & 1\%   & -0.02\% & -0.79\% & -0.06\% & -1\%  & 0\% \\
    \midrule
    \multicolumn{1}{c|}{\multirow{3}[2]{*}{E}} & FourPeople & 0.01\% & -0.02\% & 0.08\% & 3\%   & 1\%   & -     & -     & -     & -     & - \\
    \multicolumn{1}{c|}{} & Johnny & 0.03\% & 0.25\% & -0.11\% & 2\%   & 0\%   & -     & -     & -     & -     & - \\
    \multicolumn{1}{c|}{} & KristenAndSara & 0.01\% & -0.06\% & -0.01\% & 4\%   & 1\%   & -     & -     & -     & -     & - \\
    \midrule
    \multicolumn{1}{c|}{\multirow{4}[2]{*}{F}} & BasketballDrillText & 0.04\% & -0.01\% & -0.01\% & 2\%   & 1\%   & 0.03\% & -0.14\% & 0.18\% & 2\%   & -1\% \\
    \multicolumn{1}{c|}{} & ArenaOfValor & 0.02\% & -0.02\% & 0.06\% & 2\%   & 1\%   & -0.01\% & 0.02\% & -0.06\% & -2\%  & 0\% \\
    \multicolumn{1}{c|}{} & SlideEditing & 0.04\% & 0.04\% & 0.10\% & 1\%   & 1\%   & -0.07\% & -0.01\% & -0.11\% & 2\%   & 1\% \\
    \multicolumn{1}{c|}{} & SlideShow & 0.11\% & -0.07\% & -0.22\% & 1\%   & 0\%   & 0.00\% & -0.37\% & -0.39\% & 2\%   & 0\% \\
    \midrule
    \multicolumn{2}{c|}{Class A1} & 0.03\% & 0.04\% & 0.06\% & 3\%   & 1\%   & 0.01\% & -0.07\% & 0.09\% & 4\%   & 3\% \\
    \multicolumn{2}{c|}{Class A2} & 0.03\% & 0.13\% & 0.01\% & 3\%   & 2\%   & 0.00\% & -0.02\% & -0.01\% & 5\%   & 4\% \\
    \multicolumn{2}{c|}{Class B} & 0.03\% & 0.06\% & 0.07\% & 3\%   & 1\%   & 0.02\% & -0.02\% & -0.18\% & 3\%   & 1\% \\
    \multicolumn{2}{c|}{Class C} & 0.03\% & -0.05\% & -0.03\% & 1\%   & 1\%   & 0.03\% & -0.03\% & -0.18\% & 0\%   & -2\% \\
    \multicolumn{2}{c|}{Class E} & 0.01\% & 0.06\% & -0.01\% & 3\%   & 1\%   & -     & -     & -     & -     & - \\
    \midrule
    \multicolumn{2}{c|}{\textbf{Overall}} & \textbf{0.03\%} & \textbf{0.04\%} & \textbf{0.02\%} & \textbf{3\%} & \textbf{1\%} & \textbf{0.01\%} & \textbf{-0.03\%} & \textbf{-0.09\%} & \textbf{3\%} & \textbf{1\%} \\
    \midrule
    \multicolumn{2}{c|}{Class D} & 0.05\% & -0.06\% & -0.15\% & 1\%   & 1\%   & -0.01\% & -0.62\% & -0.29\% & 0\%   & 0\% \\
    \multicolumn{2}{c|}{Class F} & 0.05\% & -0.02\% & -0.02\% & 1\%   & 1\%   & -0.01\% & -0.13\% & -0.10\% & 1\%   & 0\% \\
    \bottomrule
    \end{tabular}%
  \label{RESULTS_TCQ_PRUNE}%
\end{table*}%

\subsection{Trellis Pruning}
In TCQ, RD cost calculations and examinations are performed with trellis branches in an effort to detect the optimal quantization results, which introduces high computational cost to the encoder. We propose to perform trellis pruning targeting at eliminating the unlikely quantization candidates and removing the associated transition routes. In this way, the operation complexity in BMU and ACS modules can be lowered and the total branch number $N_{TCQ}$ can be decreased. 
More specifically, the trellis pruning is carried out based on the analyses of RD cost relationships with the proposed RD model.
{The quantization candidate set $\mathcal{L}$ is composed with five candidates in principle, which can be regarded as adjusting the level of $l^{(i)}$ with different offsets $\Delta l$. Again, $l^{(i)}$ is the absolute value of the scalar quantization at position $i$, and the possible values of $\Delta l$ are as follows,}
\begin{align}
 \Delta l &\in \{-l^{(i)}, -2, -1, 0, +1\}, \text{if } l^{(i)} > 2,\nonumber \\
 \Delta l &\in \{-2, -1, 0, +1, +2\}, \text{if } l^{(i)} = 2,\nonumber \\
 \Delta l &\in \{-1, 0, +1, +2, +3\}, \text{if } l^{(i)} = 1,\nonumber \\
 \Delta l &\in \{0, +1, +2, +3, +4\}, \text{if } l^{(i)} = 0.
\end{align}
We use $l_{\Delta l}^{(i)}$ to represent the explicit quantization candidate associated to offset $\Delta l$ as follows,
\begin{align}
   l^{(i)}_{{\Delta l}} = l^{(i)} + \Delta l.
\end{align}

Supposing $C^{(i)}$ is the absolute value of the transform coefficient at position $i$, the total distortions of a coding block when quantizing $C^{(i)}$ to $l^{(i)}$ can be described as,
\begin{align}
    D_s =\sum_{k\neq i}D^{(k)} + \left(l^{(i)}\cdot Q_{step} - C^{(i)}\right)^2.
\end{align}
Analogously, if the quantization result of $C^{(i)}$ is $l^{(i)}_{{\Delta l}}$, the distortions can be expressed as follows,
\begin{align}
    D_{\Delta l} =\sum_{k\neq i}D^{(k)} + \left(l^{(i)}_{{\Delta l}} \cdot Q_{step} - C^{(i)}\right)^2.
\end{align}
The distortion difference between $l^{(i)}$ and $l_{\Delta l}^{(i)}$ with respective to $\Delta l$ can be formulated as,
\begin{align}
    \Delta D &= D_s - D_{\Delta l}  \nonumber \\
    & = -Q_{step}^2\cdot {\Delta l}^2 + 2\cdot Q_{step}\cdot(C^{(i)} - l^{(i)}\cdot Q_{step})\cdot \Delta l. 
    \label{deltaD_trellisPrunn}
\end{align}
According to Eqn.~(\ref{deltaD_trellisPrunn}), $\Delta D$ reaches the maximum value when $\Delta l$ equals to 0. Therefore, it can be noticed that $l^{(i)}$ always provides the lowest quantization distortion.

Meanwhile, the rate differences can be estimated with our proposed model in Eqn. (\ref{R_L0_L1_LP}) when $l^{(i)}$ is changed to $l^{(i)}_{\Delta l}$ as follows,
\begin{align}
    \Delta R &= R_s - R_{\Delta l}\nonumber \\
     &= \left(\alpha \cdot L_0 + \beta \cdot L_1 + \gamma \cdot R_{LP} + \epsilon\right) \nonumber \\
     &- \left[\alpha \cdot (L_0 + \eta) + \beta \cdot (L_1 + \tilde{\Delta l}) + \gamma \cdot R_{LP} + \epsilon \right] \nonumber \\
     &= -\alpha \cdot \eta - \beta \cdot \tilde{\Delta l},
     \label{deltaR_trellisPrunn}
\end{align}
where $\tilde{\Delta l}$ is the difference of the coded index when the quantization level is adjusted by $\Delta l$, and can be calculated as follows,
\begin{align}
    \tilde{\Delta l} = \tilde{l_{\Delta l}}^{(i)} - \tilde{l}^{(i)}.
\end{align}
Herein, $\tilde{l_{\Delta l}}^{(i)}$ and $\tilde{l}^{(i)}$ represent the coded indices of quantization candidates $l_{\Delta l}^{(i)}$ and $l^{(i)}$, respectively.
Typically, $\eta$ denotes the variations regarding the number of non-zero coefficients, which can be determined as follows,
\begin{equation}
    \eta = 
    \begin{cases}
    0 & \tilde{l}^{(i)} \neq 0 \text{ and } \tilde{l_{\Delta l}}^{(i)} \neq 0\\
    1 & \tilde{l}^{(i)} = 0 \text{ and } \tilde{l_{\Delta l}}^{(i)} \neq 0 \\
    -1& \tilde{l}^{(i)} \neq 0 \text{ and } \tilde{l_{\Delta l}}^{(i)} = 0. \\
    \end{cases}
\end{equation}

Subsequently, we discuss the variations of the rate and distortion with the following cases.

\subsubsection{$l^{(i)} = 0$ } in this case, $\Delta l$ and $\tilde{\Delta l}$ are both non-negative, such that $\eta$ equals to 1. Since the parameters $\alpha$ and $\beta$ are positive,
Eqn.~(\ref{deltaR_trellisPrunn}) can be written as follows,
\begin{align}
    \Delta R= -\alpha -\beta \cdot \tilde{\Delta l} \leq 0.
\end{align}
Both the distortion and rate may increase if $l^{(i)}$ is adjusted to $\tilde{l_{\Delta l}}^{(i)}$ in such scenario, which implies that the remaining quantization candidates could possibly introduce higher RD cost, leading to the coding performance loss. As such, it is eligible to directly remove the quantization candidates that are with higher levels without further calculation of the RD cost.

\subsubsection{$l^{(i)} = 1$ or $2$} 
the corresponding quantization index $\tilde{l}^{(i)}$ is ``1''. The explicit value of $\Delta R$ depends on $\tilde{\Delta l}$, which is formulated as follows,
\begin{equation}
    \Delta R = 
    \begin{cases}
        \alpha + \beta & \tilde{\Delta l} = -1\\
        -\beta \cdot \tilde{\Delta l} & \tilde{\Delta l} = 0 \text{ or } 1.
    \end{cases}
\end{equation}
It can be observed that when $\tilde{\Delta l}$ equals to -1, $\Delta R$ is a positive constant, indicating the savings of coding bits. However, it is difficult to intuitively predict the final variations of RD cost, since the associated {distortion} is also increased. Moreover, it can be inferred that positive $\tilde{\Delta l}$ leads to the increase of the coding bits. Therefore, larger quantization candidates are considered to be removed from the trellis graph in such scenarios.

\subsubsection{$l^{(i)} > 2$} Though negative $\tilde{\Delta l}$ results in the saving of the coding bits, it is still difficult to speculate the actual variation of RD cost, 
\begin{equation}
    \Delta R = 
    \begin{cases}
        \alpha + \beta \cdot \tilde{l}^{(i)} & \tilde{\Delta l} = -\tilde{l}^{(i)} \\
        -\beta \cdot \tilde{\Delta l} & \tilde{\Delta l}=-1, 0 \text{, or } 1.
        \label{pruning_Case3}
    \end{cases}
\end{equation}
For the case that $\tilde{\Delta l}$ equals to $-\tilde{l}^{(i)}$, remarkable increase of distortions could be noticed, which cannot be well remedied by the saving of coding bits. As such, it is proposed to eliminate the checking of candidate level 0.

We provide two exemplified quantization candidate sets to better illustrate the technical details of pruning. The first set $\mathcal{L}_1 = \{0[0], 1[1], 2[1], 3[2], 4[2]\}$ conforming to the former two cases, where the numbers inside and outside of the square brackets denote the quantization indices and quantization levels, respectively.
The proposed pruning procedure is demonstrated in Fig.~\ref{Fig_pruning_1}. Initially, following the transition rule of TCQ, candidates ``$4[2]$'' and ``$2[1]$'' are coupled and bounded to the quantizer $Q_0$, where State 0 and State 1 are assigned as transmitting states. Similarly, ``$1[1]$'' and ``$3[2]$'' are paired with quantizer $Q_1$ associating to State 2 and State 3. The candidate ``$0[0]$'' is specifically bounded with each state. Accumulated RD cost will be calculated for each node with the consideration of all available branches, and the one with minimal cost will be retained in the trellis graph.
With the proposed method, the unlikely-selected branches with larger quantization levels such as ``$4[2]$'' and ``$3[2]$'' are directly pruned without RD checking, leading to lower computational complexity. 
The second set $\mathcal{L}_2 = \{0[0], 2[1], 3[2], 4[2], 5[3]\}$ corresponds to the last case where transition routes of ``$0[0]$'' can be pruned, as illustrated in Fig.~\ref{Fig_pruning_2}.
Since the proposed pruning method does not have any influence on the state transition, the dequantization process remains consistent. 
In practice, we map the relationship of $l^{(i)}$ and $0$, $1$ and $2$ to the $C^{(i)}$ and associated thresholds, in order to avoid the calculation of $l^{(i)}$. In this way, given $C^{(i)}$ the pruning strategy can be efficiently determined.

\begin{table*}[htbp]
  \centering
  \scriptsize
  \caption{Performance of the combination of the proposed postponing the trellis initial point and trellis pruning method under AI and RA configurations}
    \begin{tabular}{cc|ccccc|ccccc}
    \toprule
    \multicolumn{1}{c|}{\multirow{2}[2]{*}{Class}} & \multirow{2}[2]{*}{Sequence} & \multicolumn{5}{c|}{AI}               & \multicolumn{5}{c}{RA} \\
    \multicolumn{1}{c|}{} &       & BD-Rate(Y)     & BD-Rate(U)     & BD-Rate(V)     & $TS_Q$     & $TS_{Enc}$     & BD-Rate(Y)     & BD-Rate(U)     & BD-Rate(V)     & $TS_Q$ & $TS_{Enc}$ \\
    \midrule
    \multicolumn{1}{c|}{\multirow{3}[2]{*}{A1}} & Tango2 & 0.08\% & 0.25\% & 0.38\% & 35\%  & 15\%  & 0.03\% & -0.08\% & 0.11\% & 36\%  & 5\% \\
    \multicolumn{1}{c|}{} & FoodMarket4 & 0.13\% & 0.30\% & 0.20\% & 28\%  & 11\%  & 0.09\% & 0.13\% & 0.10\% & 30\%  & 2\% \\
    \multicolumn{1}{c|}{} & Campfire & 0.07\% & 0.13\% & 0.12\% & 27\%  & 12\%  & 0.02\% & 0.17\% & 0.06\% & 27\%  & 5\% \\
    \midrule
    \multicolumn{1}{c|}{\multirow{3}[2]{*}{A2}} & CatRobot1 & 0.11\% & 0.26\% & 0.26\% & 29\%  & 13\%  & -0.01\% & -0.03\% & 0.10\% & 36\%  & 9\% \\
    \multicolumn{1}{c|}{} & DaylightRoad2 & 0.17\% & 0.41\% & 0.25\% & 28\%  & 13\%  & 0.03\% & 0.14\% & 0.08\% & 37\%  & 9\% \\
    \multicolumn{1}{c|}{} & ParkRunning3 & 0.08\% & 0.17\% & 0.23\% & 19\%  & 8\%   & 0.04\% & 0.13\% & 0.14\% & 26\%  & 5\% \\
    \midrule
    \multicolumn{1}{c|}{\multirow{5}[2]{*}{B}} & MarketPlace & 0.08\% & 0.20\% & 0.42\% & 25\%  & 11\%  & 0.03\% & -0.04\% & 0.18\% & 28\%  & 6\% \\
    \multicolumn{1}{c|}{} & RitualDance & 0.08\% & 0.23\% & 0.35\% & 22\%  & 10\%  & 0.07\% & -0.13\% & 0.16\% & 24\%  & 4\% \\
    \multicolumn{1}{c|}{} & Cactus & 0.10\% & 0.13\% & 0.30\% & 25\%  & 12\%  & 0.06\% & 0.02\% & -0.02\% & 26\%  & 6\% \\
    \multicolumn{1}{c|}{} & BasketballDrive & 0.09\% & 0.18\% & 0.26\% & 27\%  & 13\%  & 0.07\% & -0.11\% & 0.00\% & 26\%  & 3\% \\
    \multicolumn{1}{c|}{} & BQTerrace & 0.08\% & 0.12\% & 0.40\% & 20\%  & 10\%  & 0.05\% & 0.75\% & -0.12\% & 24\%  & 3\% \\
    \midrule
    \multicolumn{1}{c|}{\multirow{4}[2]{*}{C}} & BasketballDrill & 0.15\% & 0.04\% & 0.13\% & 22\%  & 11\%  & 0.08\% & -0.04\% & -0.15\% & 27\%  & 4\% \\
    \multicolumn{1}{c|}{} & BQMall & 0.10\% & 0.15\% & 0.10\% & 23\%  & 10\%  & 0.05\% & -0.04\% & -0.08\% & 23\%  & 2\% \\
    \multicolumn{1}{c|}{} & PartyScene & 0.12\% & -0.04\% & 0.30\% & 14\%  & 8\%   & 0.03\% & 0.07\% & 0.27\% & 20\%  & 3\% \\
    \multicolumn{1}{c|}{} & RaceHorses & 0.12\% & 0.02\% & 0.15\% & 17\%  & 9\%   & 0.07\% & 0.21\% & -0.32\% & 22\%  & 3\% \\
    \midrule
    \multicolumn{1}{c|}{\multirow{4}[2]{*}{D}} & BasketballPass & 0.09\% & 0.51\% & 0.16\% & 18\%  & 8\%   & 0.09\% & -0.10\% & -0.24\% & 19\%  & 5\% \\
    \multicolumn{1}{c|}{} & BQSquare & 0.16\% & -0.05\% & -0.35\% & 12\%  & 6\%   & 0.03\% & -0.18\% & -1.44\% & 17\%  & 4\% \\
    \multicolumn{1}{c|}{} & BlowingBubbles & 0.15\% & 0.16\% & 0.12\% & 13\%  & 7\%   & 0.14\% & 0.29\% & 0.39\% & 20\%  & 6\% \\
    \multicolumn{1}{c|}{} & RaceHorses & 0.12\% & 0.05\% & -0.33\% & 15\%  & 6\%   & 0.01\% & -0.33\% & 0.31\% & 17\%  & 5\% \\
    \midrule
    \multicolumn{1}{c|}{\multirow{3}[2]{*}{E}} & FourPeople & 0.12\% & 0.03\% & 0.31\% & 20\%  & 10\%  & -     & -     & -     & -     & - \\
    \multicolumn{1}{c|}{} & Johnny & 0.12\% & 0.21\% & 0.07\% & 22\%  & 10\%  & -     & -     & -     & -     & - \\
    \multicolumn{1}{c|}{} & KristenAndSara & 0.10\% & 0.00\% & 0.16\% & 21\%  & 9\%   & -     & -     & -     & -     & - \\
    \midrule
    \multicolumn{1}{c|}{\multirow{4}[2]{*}{F}} & BasketballDrillText & 0.15\% & 0.00\% & 0.05\% & 21\%  & 8\%   & 0.10\% & 0.26\% & 0.15\% & 25\%  & 7\% \\
    \multicolumn{1}{c|}{} & ArenaOfValor & 0.12\% & 0.13\% & 0.11\% & 20\%  & 6\%   & 0.08\% & 0.21\% & 0.15\% & 24\%  & 7\% \\
    \multicolumn{1}{c|}{} & SlideEditing & 0.09\% & 0.12\% & 0.22\% & 8\%   & 4\%   & 0.01\% & -0.04\% & 0.00\% & 9\%   & 2\% \\
    \multicolumn{1}{c|}{} & SlideShow & 0.14\% & 0.16\% & -0.08\% & 12\%  & 4\%   & 0.16\% & -0.30\% & -0.38\% & 10\%  & 3\% \\
    \midrule
    \multicolumn{2}{c|}{Class A1} & 0.10\% & 0.23\% & 0.23\% & 30\%  & 13\%  & 0.05\% & 0.07\% & 0.09\% & 31\%  & 4\% \\
    \multicolumn{2}{c|}{Class A2} & 0.12\% & 0.28\% & 0.25\% & 26\%  & 12\%  & 0.02\% & 0.08\% & 0.11\% & 33\%  & 8\% \\
    \multicolumn{2}{c|}{Class B} & 0.09\% & 0.17\% & 0.35\% & 24\%  & 11\%  & 0.06\% & 0.10\% & 0.04\% & 26\%  & 4\% \\
    \multicolumn{2}{c|}{Class C} & 0.12\% & 0.05\% & 0.17\% & 19\%  & 10\%  & 0.06\% & 0.05\% & -0.07\% & 23\%  & 3\% \\
    \multicolumn{2}{c|}{Class E} & 0.11\% & 0.08\% & 0.18\% & 21\%  & 10\%  &   -    &   -    &   -    &   -    & - \\
    \midrule
    \multicolumn{2}{c|}{\textbf{Overall}} & \textbf{0.11\%} & \textbf{0.16\%} & \textbf{0.24\%} & \textbf{24\%} & \textbf{11\%} & \textbf{0.05\%} & \textbf{0.08\%} & \textbf{0.03\%} & \textbf{27\%} & \textbf{5\%} \\
    \midrule
    \multicolumn{2}{c|}{Class D} & 0.13\% & 0.17\% & -0.10\% & 15\%  & 7\%   & 0.07\% & -0.08\% & -0.25\% & 18\%  & 5\% \\
    \multicolumn{2}{c|}{Class F} & 0.13\% & 0.10\% & 0.07\% & 15\%  & 6\%   & 0.09\% & 0.03\% & -0.02\% & 17\%  & 5\% \\
    \bottomrule
    \end{tabular}%
  \label{TAB_COMBIME_TEST}%
\end{table*}%

\section{Experimental Results}
\subsection{Performance Evaluations}
The proposed low complexity TCQ approaches are implemented on the VVC test platform VTM-4.0~\cite{VTM4}. Simulations are conducted conforming to the JVET Common Test Conditions (CTC)~\cite{ctc} where the recommended test sequences from class A to class F are all involved in the experiment under AI and RA configurations. The QP values are set as \{22, 27, 32, 37\}, and BD-Rates~\cite{bjontegaard2001calculation} for Y, U and V components are used to evaluate the coding performance where negative value denotes the performance gain. Computational complexity reduction is measured with the total encoding time-saving $TS_{Enc}$ and quantization time-saving $TS_{Q}$ as follows,
\begin{align}
    TS_{Enc} &= \frac{T_{Enc}^{Anc} - T_{Enc}^{Pro}}{T_{Enc}^{Anc}} \times 100\%, \nonumber \\
    TS_{Q} &= \frac{T_{Q}^{Anc} - T_{Q}^{Pro}}{T_{Q}^{Anc} } \times 100\%,
\end{align}
where $T_{Enc}^{Pro}$ and $T_{Enc}^{Anc}$ denote the total elapsed encoding time with the proposed low complexity scheme and the original anchor, respectively. Analogously, $T_{Q}^{Pro}$ and $T_{Q}^{Anc}$ stand for the quantization time of the proposed scheme and anchor, respectively.

Experimental results of the proposed scheme by postponing the trellis initial point are demonstrated in Table~\ref{TAB_ADAPTIVE} where the time-saving and the coding performance regarding each individual sequence are presented. To be more specific, the proposed method brings 22\% quantization time savings and 10\% encoding time savings under AI configuration. The performance loss is quite marginal, with 0.09\%, 0.15\% and 0.17\% BD-Rate increases for Y, U and V components, respectively. Similar performance can be observed under RA configuration where the time-saving in quantization is 25\% and the BD-Rate loss is only 0.03\%, which strikes an excellent compromise between the coding performance and computational complexity. 
Table~\ref{RESULTS_TCQ_PRUNE} shows the performance of the proposed trellis pruning method, with which 3\% quantization time reductions and 1\% encoding time reductions can be achieved. Since the branch pruning is guided by the RD model, there is only 0.03\% and 0.01\% loss on average under AI and RA configurations. Moreover, it is also interesting to see that performance gains may even happen on chroma components under the RA configuration. 
In addition, we combine the trellis pruning method with the initial point postponing method, and the experimental results are provided in Table~\ref{TAB_COMBIME_TEST}. By combining those two approaches, on average 24\% and 27\% quantization time savings can be achieved under AI and RA configurations with only 0.11\% and 0.05\% BD-Rate loss. Moreover, the overall encoding time savings are 11\% and 5\% under AI and RA configurations, respectively. 

\begin{table*}[t]
  \centering
  \scriptsize
  \caption{Performance of the proposed scheme by postponing the trellis departure point with $\mathcal{K} = 2$ (safe) under AI and RA configurations}
    \begin{tabular}{cc|ccccc|ccccc}
    \toprule
    \multicolumn{2}{c|}{\multirow{2}[2]{*}{Class}} & \multicolumn{5}{c|}{AI}               & \multicolumn{5}{c}{RA} \\
    \multicolumn{2}{c|}{} & BD-Rate(Y)     & BD-Rate(U)     & BD-Rate(V)     & $TS_Q$     & $TS_{Enc}$      & BD-Rate(Y)     & BD-Rate(U)     & BD-Rate(V)     & $TS_Q$    & $TS_{Enc}$  \\
    \midrule
    \multicolumn{2}{c|}{A1} & 0.02\% & 0.18\% & 0.07\% & 21\%  & 8\%   & 0.01\% & 0.01\% & 0.01\% & 24\%  & 3\% \\
    \multicolumn{2}{c|}{A2} & 0.01\% & 0.09\% & 0.01\% & 19\%  & 8\%   & -0.02\% & 0.08\% & 0.01\% & 26\%  & 5\% \\
    \multicolumn{2}{c|}{B} & 0.02\% & 0.00\% & 0.08\% & 16\%  & 6\%   & 0.02\% & 0.08\% & -0.17\% & 20\%  & 3\% \\
    \multicolumn{2}{c|}{C} & 0.04\% & -0.05\% & -0.04\% & 13\%  & 4\%   & 0.00\% & -0.10\% & 0.09\% & 19\%  & 2\% \\
    \multicolumn{2}{c|}{E} & 0.03\% & 0.06\% & 0.01\% & 14\%  & 5\%   & -     & -     & -     & -     & - \\
    \midrule
    \multicolumn{2}{c|}{\textbf{Overall}} & \textbf{0.02\%} & \textbf{0.05\%} & \textbf{0.03\%} & \textbf{16\%} & \textbf{6\%} & \textbf{0.01\%} & \textbf{0.02\%} & \textbf{-0.03\%} & \textbf{22\%} & \textbf{3\%} \\
    \midrule
    \multicolumn{2}{c|}{D} & 0.03\% & 0.02\% & -0.05\% & 10\%  & 3\%   & 0.00\% & -0.28\% & -0.19\% & 14\%  & 1\% \\
    \multicolumn{2}{c|}{F} & 0.05\% & -0.04\% & -0.02\% & 10\%  & 3\%   & 0.01\% & 0.00\% & 0.02\% & 13\%  & 1\% \\
    \bottomrule
    \end{tabular}%
  \label{TAB_SAFE}%
\end{table*}%

\begin{table*}[htbp]
  \centering
  \scriptsize
  \caption{Performance of the proposed scheme by postponing the trellis departure point with $\mathcal{K} = 2.5$ (risky) under AI and RA configurations}
    \begin{tabular}{cc|ccccc|ccccc}
    \toprule
    \multicolumn{2}{c|}{\multirow{2}[2]{*}{Class}} & \multicolumn{5}{c|}{AI}               & \multicolumn{5}{c}{RA} \\
    \multicolumn{2}{c|}{} & BD-Rate(Y)     & BD-Rate(U)     & BD-Rate(V)     & $TS_Q$      & $TS_{Enc}$     & BD-Rate(Y)     & BD-Rate(U)     & BD-Rate(V)     & $TS_Q$      & $TS_{Enc}$ \\
    \midrule
    \multicolumn{2}{c|}{A1} & 0.23\% & 0.81\% & 0.79\% & 36\%  & 15\%  & 0.17\% & 0.59\% & 0.49\% & 39\%  & 7\% \\
    \multicolumn{2}{c|}{A2} & 0.18\% & 0.63\% & 0.60\% & 32\%  & 14\%  & 0.13\% & 0.53\% & 0.48\% & 42\%  & 7\% \\
    \multicolumn{2}{c|}{B} & 0.21\% & 0.54\% & 0.83\% & 30\%  & 14\%  & 0.13\% & 0.70\% & 0.79\% & 36\%  & 8\% \\
    \multicolumn{2}{c|}{C} & 0.27\% & 0.45\% & 0.58\% & 24\%  & 12\%  & 0.16\% & 0.54\% & 0.60\% & 33\%  & 7\% \\
    \multicolumn{2}{c|}{E} & 0.31\% & 0.60\% & 0.61\% & 26\%  & 13\%  & -     & -     & -     & -     & - \\
    \midrule
    \multicolumn{2}{c|}{\textbf{Overall}} & \textbf{0.24\%} & \textbf{0.59\%} & \textbf{0.69\%} & \textbf{30\%} & \textbf{14\%} & \textbf{0.12\%} & \textbf{0.50\%} & \textbf{0.51\%} & \textbf{37\%} & \textbf{7\%} \\
    \midrule
    \multicolumn{2}{c|}{D} & 0.28\% & 0.52\% & 0.52\% & 20\%  & 9\%   & 0.12\% & 0.63\% & 0.22\% & 28\%  & 6\% \\
    \multicolumn{2}{c|}{F} & 0.24\% & 0.41\% & 0.47\% & 19\%  & 7\%   & 0.12\% & 0.41\% & 0.46\% & 24\%  & 5\% \\
    \bottomrule
    \end{tabular}%
  \label{TAB_K_RISK}%
\end{table*}%

\begin{table*}[htbp]
  \centering
  \scriptsize
  \caption{Complexity analyses of the original TCQ and the proposed method}
    \begin{tabular}{cccc|ccc}
    \toprule
    \multirow{2}[4]{*}{Module} & \multirow{2}[4]{*}{Branch} & \multicolumn{2}{c|}{BMU} & \multicolumn{3}{c}{ACS} \\
\cmidrule{3-7}          &       & Distortion & Rate  & Add   & Compare & Select \\
    \midrule
    TCQ   & $15 \cdot N_{m}$ &   $5\cdot N_m$    &   $5\cdot N_m$    &      $15 \cdot N_{m}$ &  $10\cdot N_m$     & $10\cdot N_m$ \\
    \midrule 
    Proposed &  \makecell[c]{$10\cdot{N_{m}^{*}}$ $\sim$$15\cdot{N_{m}^{*}}$}& $3\cdot N_m^{*} \sim 5\cdot N_m^{*}$ & $3\cdot N_m^{*} \sim 5\cdot N_m^{*}$& \makecell[c]{$10\cdot {N_{m}^{*}}$ $\sim$$15\cdot{N_{m}^{*}}$} & \makecell[c]{$5\cdot {N_{m}^{*}}$ $\sim$$10\cdot{N_{m}^{*}}$}&\makecell[c]{$5\cdot{N_{m}^{*}}$ $\sim$$10\cdot{N_{m}^{*}}$}\\
    \bottomrule
    \end{tabular}%
  \label{BMU_ACS_EXP}%
\end{table*}%
\subsection{Investigations and Discussions}
In this subsection, we provide more detailed analyses regarding the proposed schemes. First, we investigate the parameter $\mathcal{K}$ in Eqn. (\ref{K_factor}) to further evaluate the robustness of the proposed method by postponing the trellis initial point. As indicated in Eqn. (\ref{K_factor}), different settings of $\mathcal{K}$ result in varying scaling factors regarding the threshold for determining the trellis initial point. Constant settings of $\mathcal{K}$ are attempted to further verify the effectiveness of postponing the trellis initial point. More specifically, $\mathcal{K}$ is first set to a safe value as $2$. The corresponding experimental results are tabulated in Table~\ref{TAB_SAFE} where the time consumed by TCQ can be saved by 16\% and 22\% under AI and RA configurations, and the corresponding encoding time decreases are 6\% and 3\% respectively. Moreover, the coding performance loss is negligible, especially under the RA configuration. As such, the safe setting of $\mathcal{K}$ has been adopted as an encoder optimization method into VVC software~\cite{adopt-O0256}. Furthermore, we set $\mathcal{K}$ with a risky value as $2.5$ and provide the coding performance in Table~\ref{TAB_K_RISK}. It can be observed that on average 30\% quantization time is reduced under AI configuration and the encoding time-saving is 14\%. With the RA configuration, the risky $\mathcal{K}$ achieves 37\% quantization time and 7\% encoding time savings. Meanwhile, 0.24\% and 0.12\% BD-Rate losses are introduced by the risky $\mathcal{K}$ under AI and RA configurations, respectively. Therefore, different settings of $\mathcal{K}$ could bring dynamic trade-off between the complexity reduction and coding performance variation.

Subsequently, we study the proposed method from the perspective of the operation complexity which essentially relies on the quantity of stages and branches, as shown in Eqn.~(\ref{NTCQ}). 
Considering that there are three branches launched from one state node and each stage contains five states in the practical scenario, the total branch number reaches to $15 \cdot N_{m}$ for a block with $N_m$ middle stages where the start and end points are excluded for clarity. The explicit operation complexities regarding BMU and ACS modules are concluded in Table~\ref{BMU_ACS_EXP}. Supposing the proposed method by postponing the trellis initial point can reduce the number of stages from $N_m$ to $N_m^{*}$, by combining with the trellis pruning scheme, the total branch number within on coding block can be reduced to $10\cdot{N_{m}^{*}}$ at most, such that the operations in BMU and ACS modules are decreased accordingly. Therefore, the proposed method brings the overall simplification of the operation complexity in TCQ. It is also worth mentioning that quantization plays a critical role in lossy coding scenarios and alleviating the computational complexity of quantization is highly desirable for the implementation of the VVC encoder in the near future. The proposed methods are capable of achieving the progressive reductions on the quantization complexity with very marginal performance loss on the VVC based coding platform, which provides insights regarding the subsequent development of the commercial encoders.

\section{Conclusions}
This paper proposes a low complexity TCQ scheme for VVC encoding. The novelty of this paper lies in that the prominent and deterministic factors that influence the coding complexity of TCQ in VVC are identified, such that corresponding low complexity quantization schemes are developed based on theoretically established rate and distortion models.  
Experimental results show that the proposed scheme achieves 11\% and 5\% encoding time savings, and 24\% and 27\% quantization time savings on average under AI and RA configurations, respectively. The coding performance loss is very marginal where 0.11\% and 0.05\% BD-Rate increases can be observed under AI and RA configurations. Further investigations also show that the proposed method constantly reduces the operation complexity regarding the quantization and achieves progressively complexity reductions with moderate coding performance loss.

\appendices

%\section{}

%\section*{Acknowledgment}


\ifCLASSOPTIONcaptionsoff
  \newpage
\fi



\small
\bibliographystyle{IEEEtran}
\bibliography{refs}
\end{document}